\documentclass[twocolumn]{aa} 
\usepackage{graphicx}
\usepackage{amsmath}
\usepackage[]{natbib}
\usepackage{txfonts}
\usepackage{xcolor}
\def\Gaia{\textit{Gaia}\xspace}

\begin{document}

\title{News from Gaia on $\sigma$ Ori E:    a case study for the wind magnetic braking process}

\author{
H.F. Song\inst{1},
G. Meynet\inst{2,\dag},
A. Maeder\inst{2},
N. Mowlavi\inst{2},
S.R. Stroud\inst{2},
Z. Keszthelyi\inst{3},
S. Ekstr\"om\inst{2},
P. Eggenberger\inst{2},
C. Georgy\inst{2},
G. A. Wade\inst{4},
Y. Qin\inst{5,6}
}

\authorrunning{
Song et al.
}

\institute{
College of Physics, Guizhou University, Guiyang city, Guizhou Province, 550025, P.R. China
\and
Geneva Observatory, Geneva University, CH-1290 Sauverny, Switzerland
\and
Anton Pannekoek Institute for Astronomy, University of Amsterdam, Science Park 904, 1098 XH, Amsterdam, The Netherlands
\and
Department of Physics and Space Science, Royal Military College of Canada, PO Box 17000, Station Forces, Kingston, Ontario, Canada, K7K7B4
\and
Department of Physics, Anhui Normal University, Wuhu city, Anhui Province, 241000, P.R. China
\and
Center for Interdisciplinary Exploration and Research in Astrophysics (CIERA) and
Department of Physics and Astronomy, Northwestern University, Sherman Avenue, Evanston, IL 60201, USA\\
$^{\dag}$Corresponding author,
\email{georges.meynet@unige.ch}
}

  \date{Received; accepted }

 \abstract
{
$\sigma$ Ori E, a massive helium B-type star, shows a high surface rotation and a strong surface magnetic field potentially challenging the process of wind magnetic braking.
}
 {
 The Gaia satellite provides an accurate distance to that star and confirms its membership to the $\sigma$ Ori cluster. We account for these two key pieces of information to
 investigate whether single star models can reproduce the observed properties of $\sigma$ Ori E and provide new estimates for its metallicity, mass, and age.
}
 {
We compute rotating stellar models accounting for wind magnetic braking and magnetic quenching of the mass loss. Two metallicities ($Z$=0.014, with a helium mass fraction $Y$=0.273 and $Z$=0.020 with $Y=0.266$), four initial masses between 8 and 9 M$_\odot$, three initial rotations between  250 and 450 km s$^{-1}$ and three initial surface equatorial magnetic field between 3 and 7 kG are considered. Differential rotation is assumed for the internal rotation in all models.
We look for models accounting simultaneously for the observed radius, position in the HR diagram, surface velocity, and braking timescale.
 }
{
We obtain that $\sigma$ Ori E is a very young star (age less than 1 Myr) with an initial mass around 9 M$_\odot$, a surface equatorial magnetic field around 7 kG and having a metallicity $Z$ (mass fraction of heavy elements) around 0.020. No solution is obtained with the present models for a metallicity $Z$=0.014. The initial rotation of the models fitting $\sigma$ Ori E  is not much constrained and can be anywhere in the range studied here.
Because of its very young age, models predict no observable changes of the surface abundances due to rotational mixing.
}
{The simultaneous high surface rotation and high surface magnetic field of $\sigma$ Ori E  may simply be a consequence of its young age. This young age implies that the processes responsible for producing the chemical inhomogeneities that are observed at its surface should be rapid. Thus for explaining the properties of $\sigma$ Ori E, there is no necessity to invoke a merging event although
such a scenario cannot be discarded.
Other stars (HR 5907, HR 7355, HR 345439, HD 2347, CPD -50$^{o}3509$) showing similar properties as $\sigma$ Ori E  (fast rotation and strong surface magnetic field) may also be very young stars, although
determination of the braking timescales is needed  to confirm such a conclusion.}

\keywords{stars:rotation; stars: abundances; stars: magnetic field; stars:evolution}

\maketitle
%

\section{Introduction}

The star $\sigma$ Ori E (HD 37479) is a magnetic, He-strong, Main-Sequence star of spectral type B2Vpe \citep[][]{Osmer1974, Landstreet1978}. It shows variations of helium abundance across its surface \citep[][]{Krticka2020}. It has a strong surface magnetic field whose morphology combines a dipolar component, with a polar strength $B_{\rm dip}$ = 7.3 - 7.8 kG with obliquity 47$^\circ$-59$^\circ$, and a smaller
non-axisymmetric quadrupole component with strength $B_{\rm qua}$=3 - 5 kG \citep{Oksala2015}. It has a relatively fast rotation \citep[P$_{\rm rot}$ is equal to 1.19 days, implying, see below,  a surface equatorial velocity of $\sim$160 km s$^{-1}$,][]{Townsend2010}. 

Direct measurements of rotational period change exist for just four magnetic stars: CU Vir, HD 37776, $\sigma$ Ori E, and HD 142990. The rotation of $\sigma$ Ori E is observed to slow down at approximately the rate predicted by analytical prescriptions of magnetic braking \citep{Ud-Doula2002,Townsend2010,Oksala2012}. Interestingly, in the three other cases, apparently cyclical period changes - including episodes of rotational acceleration - have been observed  \citep{Mikul2011,Shultz2019b}.

The first measure  of period change for $\sigma$ Ori E was done by \citet{Miku2008}. These authors have obtained for the ratio $P_{\rm rot}/\dot{P}$, where
$P_{\rm rot}$ is the rotation period and $\dot{P}$, the rate of its change, a value of 0.25 Myr. We shall call this ratio the present-day braking timescale. For this ratio, \citet{Townsend2010} gives a value of
1.3 Myr, so considerably longer. Still more recently, Petit et al (in preparation) obtain a value of 1.12 Myr\footnote{New $\dot{P}_{\rm rot}$ of 91.9 ms per year is measured by Petit et al. (in prep). This gives, considering a rotation period of 1.19 days, a value of the ratio equal to 1.12 Myr.}. In the present paper, we shall stick to the published value by \citet{Townsend2010}.
The present braking timescale is not giving an estimate of the age of the star or of
the duration of the period during which the star has been braked because we do not know how $P_{\rm rot}/\dot{P}$ evolves.
For instance, the braking timescale can be significantly shorter than the age of the star if, for instance, the surface is continuously accelerated by a transport of angular momentum 
from the core to the envelope, or if the star has been spun up by an interaction or a merging with a companion. The braking timescale can also be larger than the age
of the star in case the star would be in a very early phase of its evolution.

In this paper we want to identify what is the most probable status of $\sigma$ Ori E. Is it a star at a very early stage of its evolution, in which case
there will no problem to explain both its strong surface magnetic field and rapid rotation. Is it an evolved star, with a braking timescale shorter than its age? In that case, either
some efficient internal transport of angular momentum would be required for single star to reproduce the observed properties or an interaction 
with a close companion has to be invoked.

If the star $\sigma$ Ori E belongs to the $\sigma$ Ori cluster as has been assumed by \citet[][]{Townsend2013}, and if the age of this cluster
is around 2-3 Myr as given by \citet{Sherry2008} and \citet{Caba2007}, then we would have a configuration where the age of the star is slightly
larger than the present braking timescale. If $\sigma$ Ori E indeed belongs to the  $\sigma$ Ori cluster and the determined age is correct 
(actually age determinations are very model dependent), we are left with one of the two last possibilities, either a single star with some internal angular momentum transport
or a binary with, at a given time, an interaction that has spun up the star (by tides, mass accretion or merging process).

In the present paper, we want first take profit from the Gaia Data Release 2 to obtain a better estimate of the distance of $\sigma$ Ori E 
and get some clues about its belonging or not to the $\sigma$ Ori cluster. As explained below, knowing the distance allows the determination of the radius of the star
(thanks to a photometric estimate of the angular diameter). From the observed rotation period and the stellar radius, one can obtain the surface velocity. From the radius and the effective temperature, one can estimate  the luminosity of $\sigma$ Ori E. Adding the observed value of $P_{\rm rot}/\dot{P}$, this makes five constraints (radius, surface velocity, luminosity, effective temperature and present braking timescale) that any model has to reproduce. The question that we want to address here: do single star models exist able to reproduce these five constraints? If yes, what would be the range of ages for
$\sigma$ Ori E? What would be other, still non observed properties that such models would predict for $\sigma$ Ori E?

We present in section 2 the observed properties of $\sigma$ Ori E.
The physics of our stellar models is explained in Sect. 3.  Section 4 looks for models reproducing the above five observed constraints.
The results are discussed in Sect.~5 and the main conclusions are given in Sect.~6.

\section{Observational constraints of $\sigma$ Ori E}

\subsection{Distance of $\sigma$ Ori E}
\label{Sect:Gaia}

Distance estimates for $\sigma$ Ori E before the \Gaia data releases relied on the distance to the cluster.
This distance ranges in the literature from $352^{+166}_{-168}$~pc to $473\pm 33$~pc \citep[see Table 2 in][]{Caballero18NM}.
It is only with the parallaxes obtained by \Gaia that an individual distance to $\sigma$ Ori E has been provided.
We note that no parallax was provided by Hipparcos for this star.

\begin{figure}
	\centering
	\includegraphics[width=9cm]{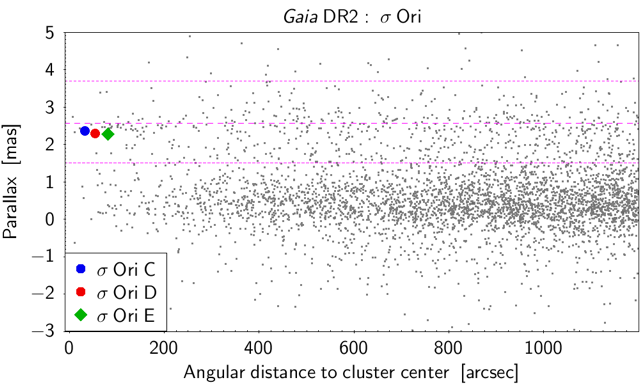}
	\caption{\Gaia DR2 parallaxes of all stars in the direction of the $\sigma$~Ori cluster versus their \Gaia DR2 distance to the cluster center taken at (RA,~Dec) = (84.675,~$-$2.6)~deg from Simbad.
	         The dashed and dotted horizontal lines indicate the mean and one standard deviation limits, respectively, of the parallaxes as determined by \cite{Caballero18} from cluster members.
	         The positions of $\sigma$~Ori~C, D and E in the diagram are indicated by the blue, red filled circles and the green filled diamond, respectively.
	}
	\label{Fig:parallax}
\end{figure}
\begin{figure}
	\centering
	\includegraphics[width=8cm]{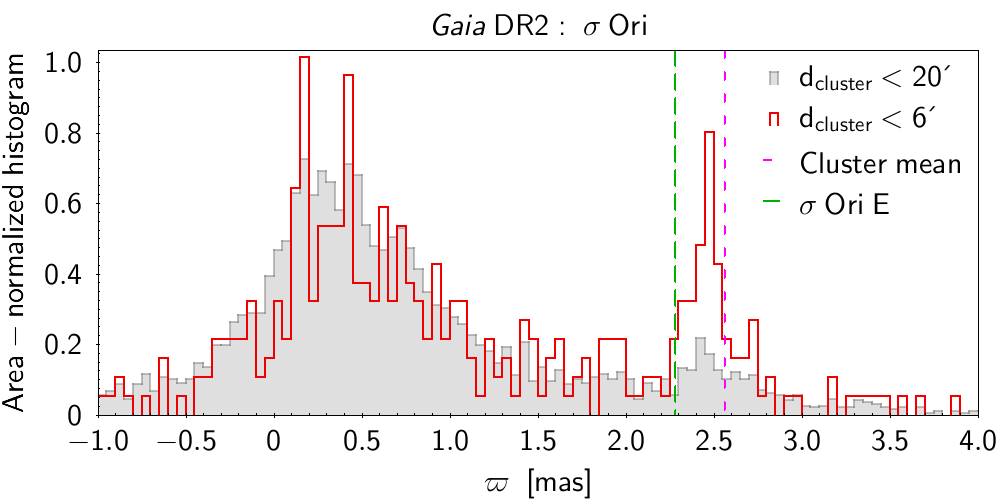}
	\caption{Histograms of the \Gaia DR2 parallaxes of all stars within a distance on the sky of 6 arcmin (open red histogram) and 20 arcmin (filled gray histogram) to the cluster center taken at the same position as in Fig.~\ref{Fig:parallax}.
	         The histograms are area-normalized.
	         The vertical short magenta dashed line locates the mean \Gaia DR2 parallax of the cluster as determined by \cite{Caballero18}.
	         The vertical long green dashed line locates the \Gaia DR2 parallax of $\sigma$~Ori~E.
	}
	\label{Fig:histo_parallax}
\end{figure}

The proper motion of $\sigma$~Ori~E confirms that this star is a member of the young $\sigma$~Orionis open cluster.
Based on a sample of 281 star members including brown dwarfs, \cite{Caballero18} derive a mean parallax of the cluster from \Gaia DR2 parallaxes of $2.56\pm 0.29$~mas.
This corresponds to a mean distance of $391^{+50}_{-40}$~pc.
The cluster is very extended in space, making it unprecise to assume the distance of an individual member to be equal to the distance of the cluster.
This is shown in Fig.~\ref{Fig:parallax}, which plots the \Gaia DR2 parallaxes versus sky distance of all stars within 30~arcmin from the center of the cluster.
The cluster stands out from background stars at parallaxes within the $2.56\pm0.29$~mas limits, shown by the magenta dotted lines in the figure.
It is even more clearly seen in Fig.~\ref{Fig:histo_parallax}, which plots in red the histogram of the parallaxes of stars within 6~arcmin from the cluster center against the histogram in filled gray of stars within 20~arcmin from the cluster center.


\begin{figure}
	\centering
	\includegraphics[width=8cm]{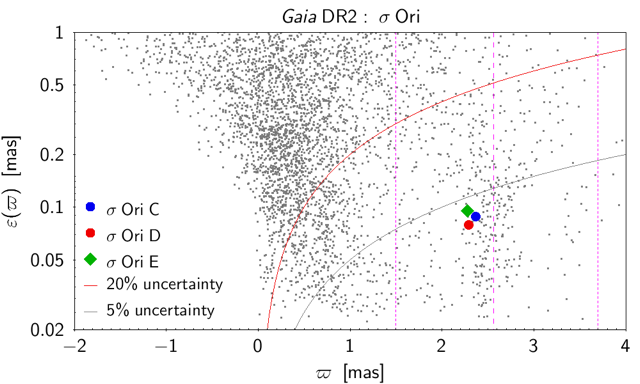}
	\caption{\Gaia DR2 parallax uncertainties versus parallax of all stars within a distance on the sky of 20 arcmin to the cluster center taken at the same position as in Fig.~\ref{Fig:parallax}.
	         The dashed and dotted vertical lines indicate the mean and one standard deviation limits, respectively, of the parallaxes as determined by \cite{Caballero18} from cluster members.
	         The red and gray solid lines indicate the parallax uncertainty limit at any given parallax below which the relative uncertainty is less than 20\% and 5\%, respectively.
	         The positions of $\sigma$~Ori~C, D and E in the diagram are indicated by the blue, red filled circles and the green filled diamond, respectively.
	}
	\label{Fig:parallaxError}
\end{figure}

\begin{figure*}
   \centering
      \includegraphics[width=13.cm]{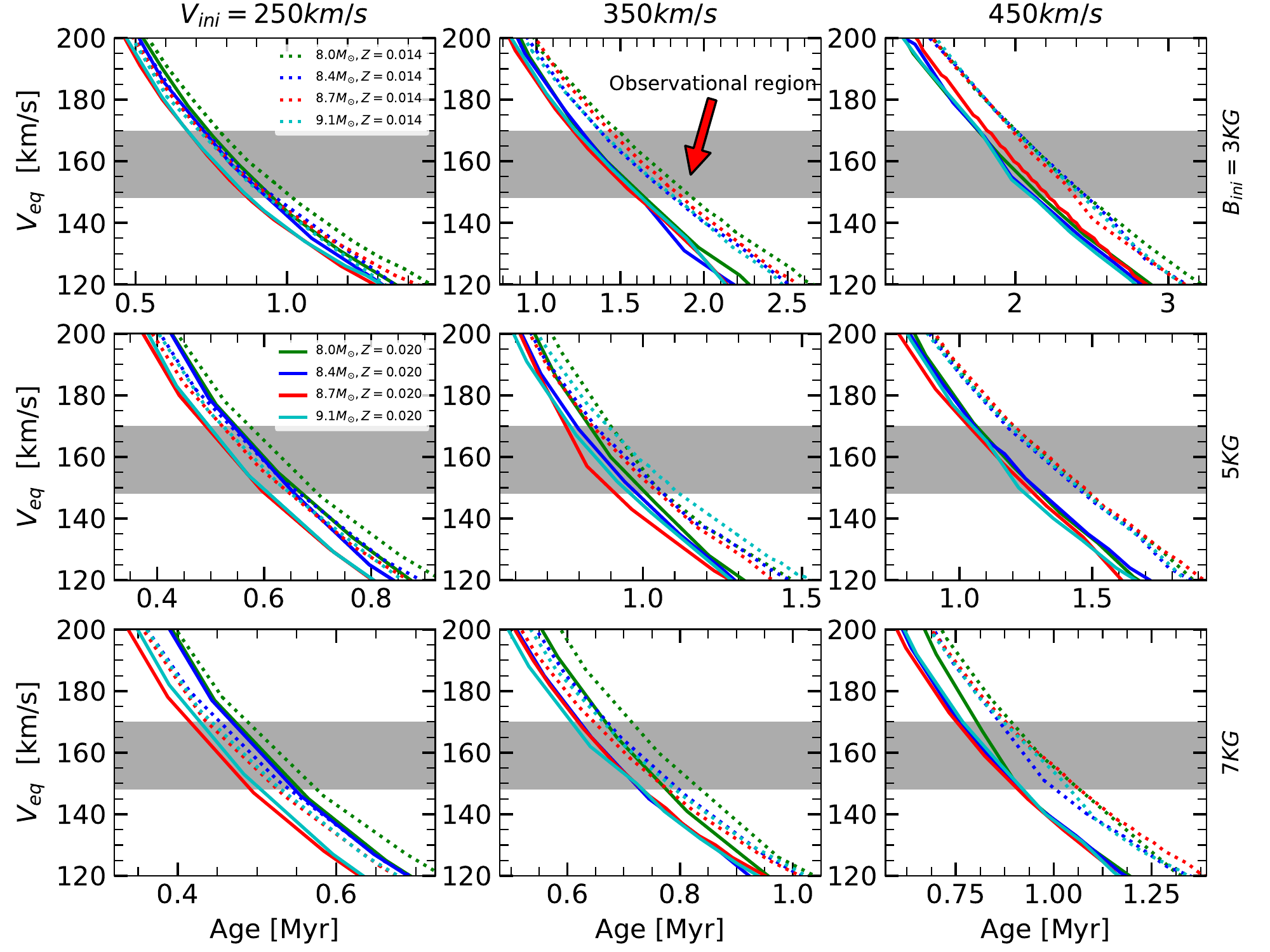} 
       \includegraphics[width=13.cm]{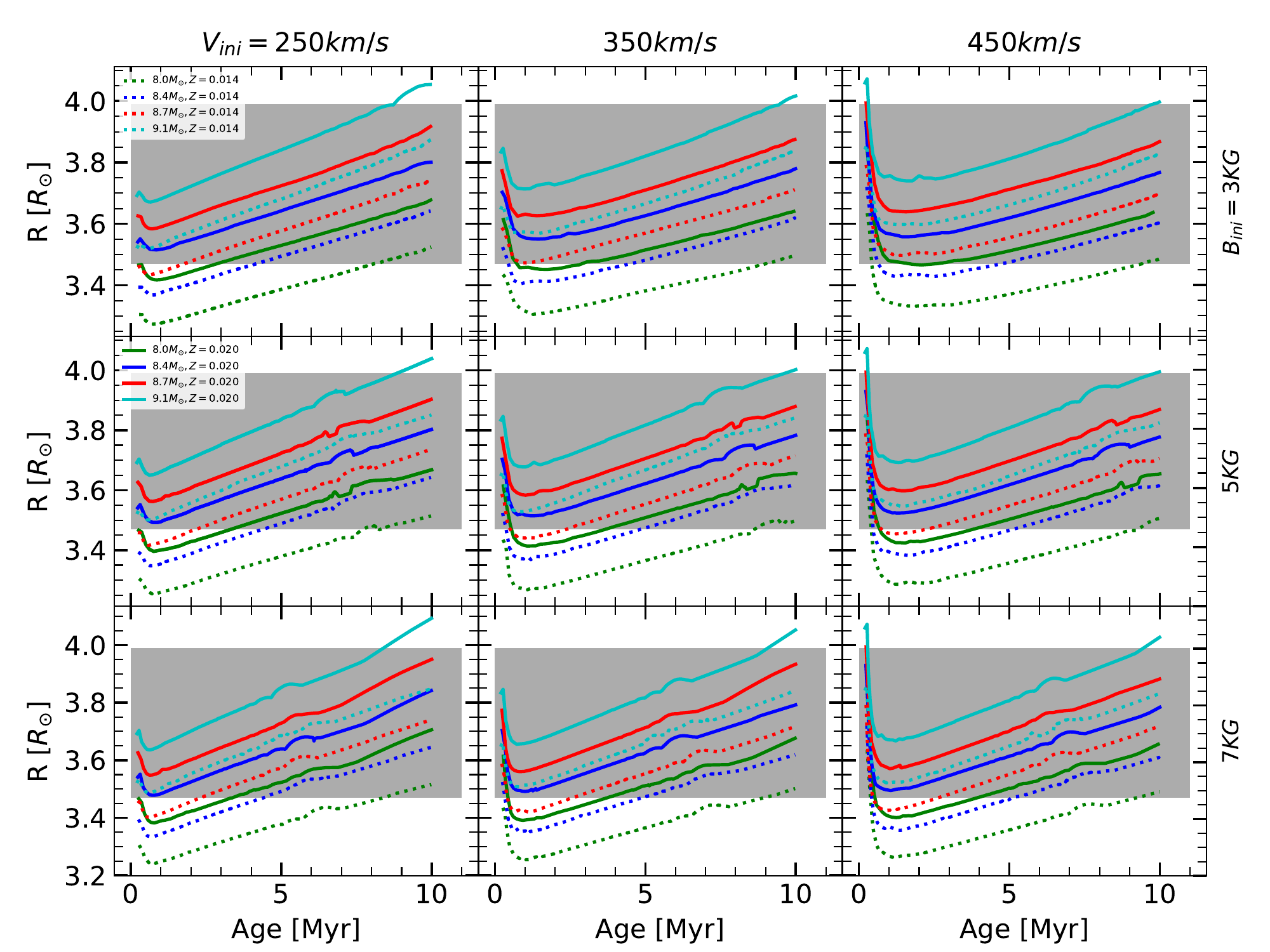}
      \caption{{\it Top panel:} Evolution of the surface equatorial velocity as a function of the age of the star for stellar models with various initial masses, metallicities, rotations, and surface
      magnetic fields. The shaded areas show the range of values for the surface equatorial velocity that can be deduced from the observed rotational period and from the stellar radius
      determined from the angular diameter and the Gaia distance. {\it Bottom panel:} Evolution of the stellar radius as a function of the age of the star for stellar models with various initial masses, metallicities, rotations, and surface magnetic fields. The light shaded area show the range of values for the stellar radii that can be deduced from  from the angular diameter and the Gaia distance.
}
   \label{veqtime}
   \end{figure*}

\begin{figure*}[h!]
   \centering
     \includegraphics[width=13.cm]{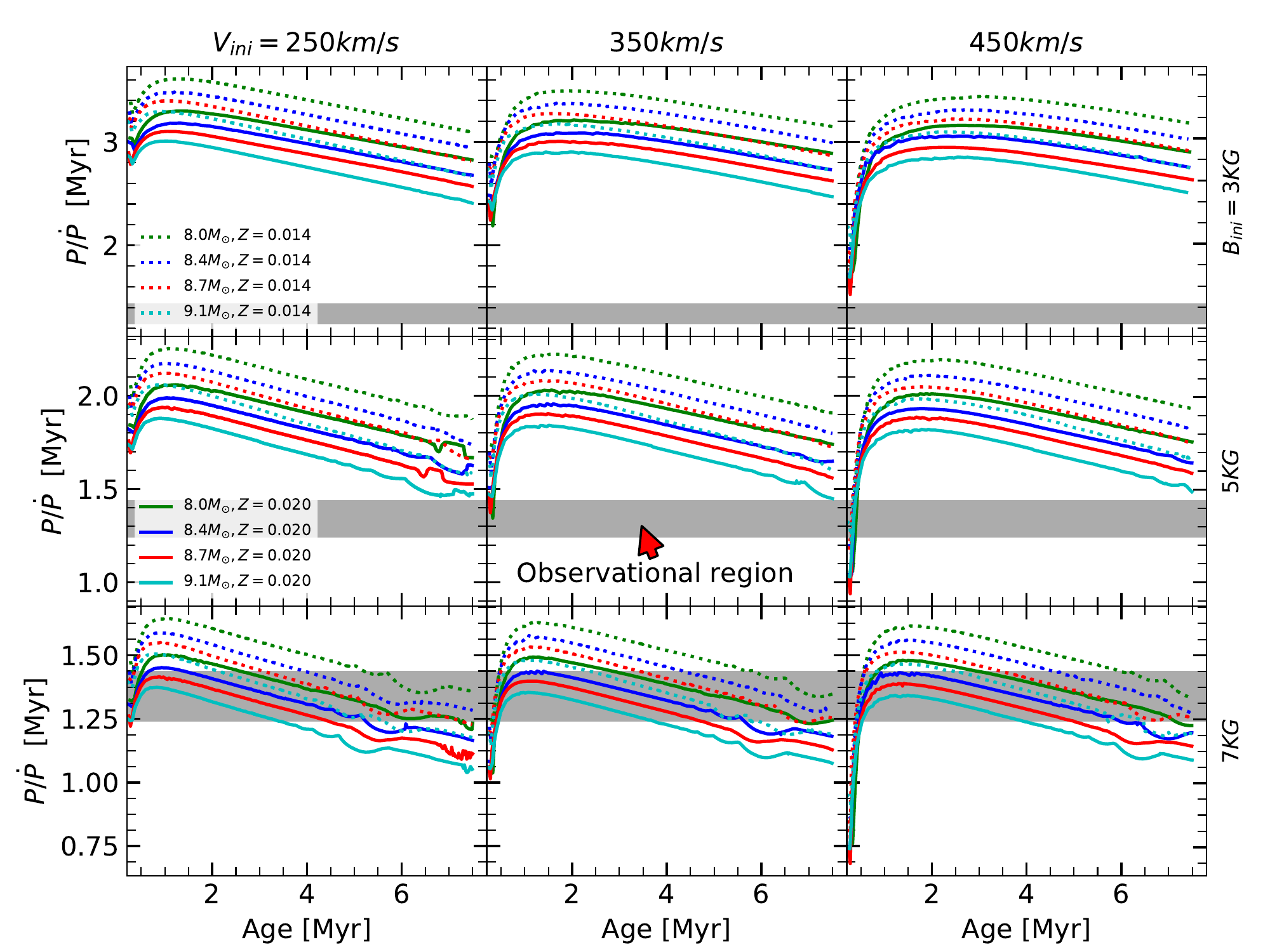}
     \includegraphics[width=13.cm]{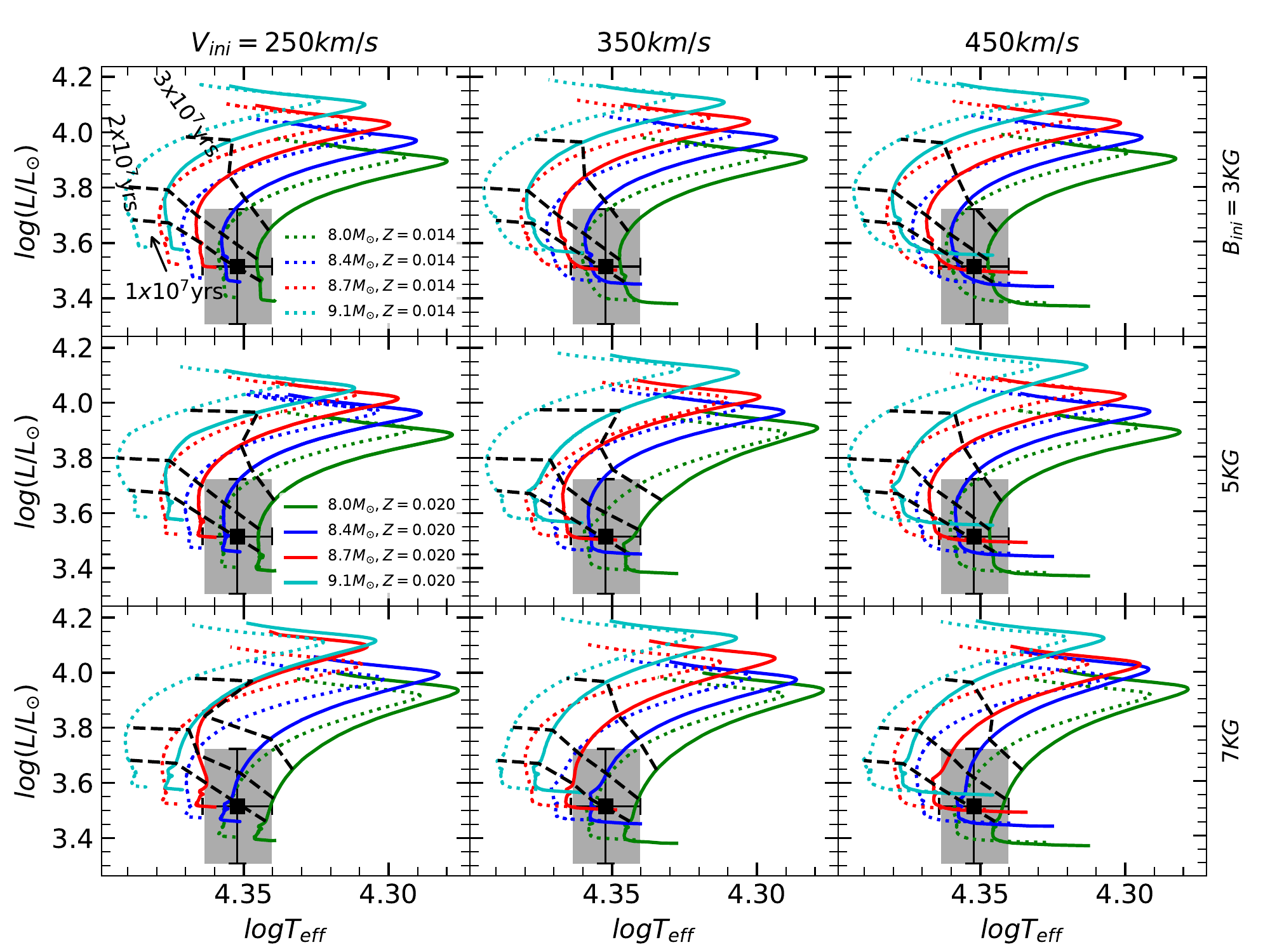}
      \caption{{\it Top panel:} Evolution of $\dot{P}/P$, where $P$ is the rotation period, for stellar models with various initial masses, metallicities, rotations, and surface
      magnetic fields. The light shaded area show the range of values for the braking timescale as deduced by \citet{Townsend2010}. {\it Bottom panel:}
 Evolutionary tracks for stellar models with various initial masses, metallicities, rotations, and surface
      magnetic fields. Three black dashed lines indicate the isochrones with ages ranging from 10 Myr to 30 Myr. The position of $\sigma$ Ori E is indicated.
}
   \label{pdotptime}
   \end{figure*}

\begin{table}[h]
\scriptsize
\begin{center}
\caption{Observational data on  $\sigma$ Ori E}
\begin{tabular}{lcl}
\hline\hline
& & \\
\multicolumn{3}{c}{\bf Gaia Astrometry} \\
& & \\
Distance  &  $439^{+20}_{-19}$ pc &  \citet{DR22018} \\
  %
%
& & \\
\hline
& & \\
\multicolumn{3}{c}{\bf Spectropolarimetry} \\
&  &\\
Polar magnetic field, $B_{\rm p}$   &  7.3-7.8 kG  & \citet{Oksala2012, Oksala2015} \\
     & & \citet{Townsend2010,Townsend2013} \\
$\log g$               & 4.0$\pm$0.5  & \citet{Oksala2012, Oksala2015}\\
%
& & \\
\hline
& & \\
\multicolumn{3}{c}{\bf Photometry} \\
&  &\\
Rotational period, $P_{\rm rot}$  &  1.19 days  & \citet{Townsend2010,Townsend2013} \\
$\dot{P}_{\rm rot}$  & 77 mms per year  & \citet{Townsend2010,Townsend2013}\\
Angular diameter & 0.079$\pm$0.002 mas & \citet{GH1982}\\
$\log T_{\rm eff} $ & 4.352$\pm$0.012 & \citet{GH1982}\\
  & &   \citet{Reiners2000}\\
& & \\
\hline
& & \\
\multicolumn{3}{c}{\bf Spectroscopy} \\
 & & \\
$\log g$               & 4.2$\pm$0.2  & \citet{Shultz2019a}\\
$\log T_{\rm eff}$ & 4.362$\pm$0.04 & \citet{Hunger1989}\\ 
Log L/L$_\odot$ &   3.50$\pm$0.20 &  \citet{Shultz2019a}\\         
$\upsilon\sin i$  & 140$\pm$10 km s$^{-1}$ & \citet{Oksala2012, Oksala2015} \\
$\epsilon_{\rm C}=\log{N_{\rm C} \over N_{\rm H}}$ & -4.5$\pm$0.5 dex & \citet{Oksala2012, Oksala2015} \\
$\epsilon_{\rm He}=\log{N_{\rm He} \over N_{\rm H}}$ & -0.85$\pm$0.25 dex & \citet{Oksala2012, Oksala2015} \\
& & \\
\hline\hline
\end{tabular}
\label{tableobs}
\end{center}
\end{table}

\begin{table}[h]
\begin{center}
\scriptsize
\caption{Properties of $\sigma$ Ori E to be fitted by stellar models. }
\begin{tabular}{cc}
\hline\hline
  & \\
Radius.               & 3.73$\pm$0.26 R$_\odot$  \\
Surface equatorial velocity  &  159 $\pm$ 11 km s$^{-1}$\\
$P/\dot{P}$          &   1.34$\pm$ 0.10 My          \\
Log L/L$_\odot$ &   3.50$\pm$0.19            \\
Log T$_{\rm eff}$&   4.352$\pm$0.012       \\
  & \\
\hline\hline
\end{tabular}
\label{tableconst}
\end{center}
\end{table}

The three bright stars $\sigma$~Ori~C, D and E are highlighted in Fig.~\ref{Fig:parallax} with colored filled circles.
They are close to the center of the cluster and have parallaxes smaller than the mean of the cluster.
$\sigma$~Ori~E, in particular, at a \Gaia DR2 angular distance of 80.4~arcsec on the sky from the cluster center has a parallax of $2.28\pm0.10$~mas, i.e. a distance of $439^{+20}_{-19}$~pc.
This places it towards the furthermost edge of the cluster (like for $\sigma$~Ori C and D), as shown by the long-dashed vertical line in Fig.~\ref{Fig:histo_parallax}.
This comes in contrast to the distance estimate of $387.5 \pm 1.3$~pc for the triple $\sigma$~Ori~A and B system determined by \cite{SchaeferHummelGies_etal16} from interferometric measurements (no \Gaia DR2 parallax is available for $\sigma$~Ori~AB as it is too bright).
The various astrometric quality checks are nevertheless good for $\sigma$~Ori~E: ten visibility periods are used in the astrometric calculation, the astrometric excess noise of 0.25~mas is not significant%
, and the parallax uncertainty is less than 5\% (see below). 
The distance to $\sigma$~Ori~E derived from the \Gaia DR2 parallaxes should thus be reliable.

The parallax uncertainties are shown in Fig.~\ref{Fig:parallaxError} versus parallaxes for all stars within a sky distance of 20~arcmin of the cluster center.
The great majority of stars with parallaxes between 2~mas and 3~mas, i.e. potential cluster members, have a parallax uncertainty better than 20\%, and almost half of them, including $\sigma$~Ori~E, have parallax uncertainties better than 5\%.

An estimate of the systematic parallax uncertainties is more difficult to evaluate, as it depends on many factors \citep[see][]{Luri_etal18}.
If we take the systematic uncertainty of 0.029~mas derived by \citet{Luri_etal18} from the distribution of quasars, the corrected distance to $\sigma$~Ori~E would be 433~pc, which is within the uncertainties of the uncorrected distance. 

With respect to the DR2, the early DR3 (EDR3) still allows a slight improvement of the parallax.
It gives a parallax of $2.31\pm 0.06$~mas implying a distance of $433\pm 11$~pc.
This EDR3 distance is good agreement with the distance of 439~pc obtained above from DR2.

In addition, new photometric attributes published in EDR3 from the Image Parameters Determination (IPD) module \citep{Lindegren_etal20} confirm the absence of significant flux structures in the image window around $\sigma$~Ori~E.
A first parameter, \texttt{ipd\_frac\_multi\_peak}, indicates the fraction of valid transits for which another peak is observed in the image window around the source.
For $\sigma$~Ori~E, this parameter equals zero, which means that only one peak is detected in the image window of $\sigma$~Ori~E for all transits used in the astrometric solution.
A second parameter, \texttt{ipd\_gof\_harmonic\_amplitude}, with a small value of 0.015 for $\sigma$~Ori~E, indicates that the goodness-of-fit of the flux distribution is independent of scan angle.
Both these indicators point to the absence of significant structures of the flux distribution around $\sigma$~Ori~E such as could result from the presence of a second source or of specific patch patterns around the source, and thereby support the reliability of the distances derived from DR2 and EDR3 parallaxes.

\subsection{Observed properties of $\sigma$ Ori E}

\begin{table*}[h]
\begin{center}
\scriptsize{
\caption{For each stellar model characterized by an initial mass, metallicity, rotation and surface magnetic field is indicated the age range where the model
fits the observed surface velocity, the observed radius and the observed braking timescale. The column labeled by HRD indicates whether the model fits (Y) or not (N) the observed position in the HR diagram.
Red colors indicate age ranges and HRD positions that are not compatible with the the age range given by the surface velocity constraint. Only seven models highlighted with bold face satisfy all the constraints considered in this table.}
\begin{tabular}{ccc|cccc|cccc}
\hline\hline
  & & & & & & & &  & &\\
$M_{\rm ini}$ & $\upsilon_{\rm ini}$ & $B_{\rm eq., ini}$  & \multicolumn{4}{c|}{Age ranges in Myr deduced from}  & \multicolumn{4}{|c}{Age ranges in Myr deduced from} \\
$M_\odot$  &  ${\rm km} \over {\rm s}$ & kG &  $\upsilon_{\rm eq}$ & R & $P\over \dot{P}$ &  HRD & $\upsilon_{\rm eq}$ & R & $P \over \dot{P}$ & HRD  \\
 & & & & & &  & &  & &\\
  & & &   \multicolumn{4}{c|}{Z=0.014} &  \multicolumn{4}{|c}{Z=0.020} \\
8.0  & 250 & 3 & 0.81-1.02 & \textcolor{red}{> 8.20} & \textcolor{red}{No Sol.}&                        Y     & 0.74-0.95  & \textcolor{red}{>3.0}   &   \textcolor{red}{No Sol.}  & Y  \\
8.4  & 250 & 3 & 0.72-0.96 & \textcolor{red}{> 4.25} & \textcolor{red}{No Sol.}& \textcolor{red}{N}   & 0.73-0.94  & >0.0                            &  \textcolor{red}{No Sol.}   & Y \\
8.7  & 250 & 3 & 0.72-0.94 &  \textcolor{red}{>1.70} & \textcolor{red}{No Sol.}& \textcolor{red}{N}   & 0.67-0.87  & >0.0.                           &  \textcolor{red}{No Sol.}   & Y \\
9.1  & 250 & 3 & 0.71-0.94 &  > 0.0                           & \textcolor{red}{No Sol.}& \textcolor{red}{N}   & 0.67-0.88 & 0.00-8.7                       &  \textcolor{red}{No Sol.}   &  \textcolor{red}{N}\\
  & & & & & &  & &  & &\\
8.0  & 250 & 5 & 0.56-0.70 & \textcolor{red}{> 7.8}     &\textcolor{red}{No Sol.}&                          Y  & 0.55-0.67 &  \textcolor{red}{>3.3}    &\textcolor{red}{No Sol.}& Y  \\
8.4  & 250 & 5 & 0.53-0.67 & \textcolor{red}{>4.5}      &\textcolor{red}{No Sol.}&  \textcolor{red}{N} & 0.54-0.66 &  >0.0                             &\textcolor{red}{No Sol.}& Y \\
8.7  & 250 & 5 & 0.52-0.65 & \textcolor{red}{>2.0}      &\textcolor{red}{No Sol.}&  \textcolor{red}{N} & 0.48-0.60 &  >0.0                             &\textcolor{red}{No Sol.}& Y \\
9.1  & 250 & 5 & 0.52-0.65 & >0.0                               &\textcolor{red}{No Sol.}&  \textcolor{red}{N} & 0.50-0.61 & 0.00-8.80                     &\textcolor{red}{No Sol.}&  \textcolor{red}{N} \\
  & & & & & &  & &  & & \\
8.0  & 250 & 7 & 0.49-0.57 & \textcolor{red}{>8.4}      & \textcolor{red}{>5.2}.            &                          Y & 0.47-0.55        & \textcolor{red}{>3.5} & <0.5 or 2.6-7.3 & Y   \\
8.4  & 250 & 7 & 0.45-0.54 &  \textcolor{red}{>4.5}     & \textcolor{red}{<0.3 or >3.9} &\textcolor{red}{N} & {\bf 0.47-0.55} & >0.0                          & <0.7 or 1.4-5.3 & Y  \\
8.7  & 250 & 7 & 0.43-0.53 &  \textcolor{red}{>2.3}     &  \textcolor{red}{<0.4 or >3.} &\textcolor{red}{N}  & {\bf 0.42-0.49} & >0.0                          & <4.5                 & Y  \\
9.1  & 250 & 7 & 0.44-0.53 &  >0.0                              &  \textcolor{red}{<0.4 or >2.1} &\textcolor{red}{N}& 0.43-0.51        & <8.40                        & <3.4.                  & \textcolor{red}{N}\\
  & & & & & &  & & & & \\
8.0  & 350 & 3 & 1.48-1.92 & \textcolor{red}{>9.3} & \textcolor{red}{No Sol.} &                          Y & 1.25-1.65 &  \textcolor{red}{<0.7or >2.8}         &\textcolor{red}{No Sol.}        & Y  \\
8.4  & 350 & 3 & 1.38-1.81 & \textcolor{red}{>4.6} & \textcolor{red}{No Sol.} &                          Y& 1.25-1.63  &  >0.0                                             &\textcolor{red}{No Sol.}        & Y \\
8.7  & 350 & 3 & 1.44-1.86 & >0.0                         & \textcolor{red}{No Sol.} &                          Y & 1.20-1.60  &  >0.0                                            &\textcolor{red}{No Sol.}        & Y \\
9.1  & 350 & 3 & 1.38-1.81 & >0.0                         & \textcolor{red}{No Sol.} &  \textcolor{red}{N} & 1.23-1.63 &  <9.5                                            &\textcolor{red}{No Sol.}         & Y\\
  & & & & & &  & &  & &\\
8.0  & 350 & 5 & 0.90-1.06 & \textcolor{red}{>8.7}&  \textcolor{red}{No Sol.} &                       Y  & 0.83-1.02        & \textcolor{red}{>4.7}                  &\textcolor{red}{<0.3 or > 7.5}& Y  \\
8.4  & 350 & 5 & 0.85-1.06 & \textcolor{red}{>4.8}& \textcolor{red}{No Sol.}&                         Y & 0.79-0.98        & >0.0                                            &\textcolor{red}{<0.3 or > 7.5}& Y \\
8.7  & 350 & 5 & 0.84-1.04 & \textcolor{red}{>2.5}& \textcolor{red}{No Sol.}& \textcolor{red}{N}& 0.75-0.91.       & >0.0                                            &\textcolor{red}{<0.3 or > 7.5}& Y \\
9.1  & 350 & 5 & 0.89-1.04 & >0.0                        & \textcolor{red}{No Sol.}& \textcolor{red}{N} & 0.77-0.96.       & <9.5                                            &\textcolor{red}{No Sol.}        &Y\\
  & & & & & &  & &  & & \\
8.0  & 350 & 7 & 0.72-0.83 & \textcolor{red}{>9.1} &\textcolor{red}{<0.5 or >5.9}    &                          Y & 0.67-0.77         & \textcolor{red}{<0.5 or >3.9}  & \textcolor{red}{0.3-0.7 or > 2.8}         & Y \\
8.4  & 350 & 7 & 0.67-0.80 & \textcolor{red}{>4.9} &\textcolor{red}{0.3-0.5 or >4.3}&                         Y &  {\bf 0.62-0.73} & >0.0                                       & 0.3-5.7.                                              & Y \\
8.7  & 350 & 7 & 0.65-0.77 & \textcolor{red}{>2.5} &\textcolor{red}{0.3-0.5 or >3.3}& \textcolor{red}{N} & {\bf 0.62-0.73} & >0.0                                        & 0.4-4.7                                               & Y \\
9.1  & 350 & 7 & 0.67-0.78 & >0.0                          &\textcolor{red}{0.3-0.6 or >2.3}& \textcolor{red}{N} & {\bf 0.61-0.73} & 0.0-9.0                                   & 0.5-3.7                                               & Y\\
  & & & & & &  & & & & \\
8.0  & 450 & 3 & 2.00-2.50 &  \textcolor{red}{0.00-0.50 or >9.5} &\textcolor{red}{No Sol.}  &                        Y  & 1.76-2.17 & >0.0                                             &\textcolor{red}{No Sol.} & Y  \\
8.4  & 450 & 3 & 2.00-2.50 &  \textcolor{red}{0.00-0.60  or >4.8}&\textcolor{red}{No Sol.}  &                        Y  & 1.76-2.15 & >0.0                                             &\textcolor{red}{No Sol.} & Y \\
8.7  & 450 & 3 & 1.98-2.40 &  > 0.0                                             &\textcolor{red}{No Sol.}  &                        Y  & 1.81-2.13 & >0.25                                            &\textcolor{red}{No Sol.}   & Y \\
9.1  & 450 & 3 & 1.99-2.45 & > 0.0                                              &\textcolor{red}{No Sol.}  &\textcolor{red}{N}  & 1.76-2.23 & 0.35 - 9.6                                     &\textcolor{red}{No Sol.}  & Y\\
  & & & & & &  & &  & &\\
8.0  & 450 & 5 & 1.19-1.49 & \textcolor{red}{0.00-0.40 or >9.2}  &\textcolor{red}{0.3-0.4}   &                        Y & 1.06-1.36 &  \textcolor{red}{<0.7 or >3.9}.        &\textcolor{red}{0.35-0.50}        & Y  \\
8.4  & 450 & 5 & 1.17-1.47 & \textcolor{red}{0.00-0.50  or >5.2} & \textcolor{red}{0.3-0.4}  &                       Y  & 1.06-1.37 &  >0.0                                              &\textcolor{red}{0.35-0.50}        & Y \\
8.7  & 450 & 5 & 1.19-1.49 & \textcolor{red}{0.00-0.70  or >2.3} & \textcolor{red}{0.3-0.4}  &                       Y  & 1.03-1-32 &  >0.2                                              &\textcolor{red}{0.35-0.50}       & Y \\
9.1  & 450 & 5 & 1.18-1.47 & >0.0                                               &\textcolor{red}{0.0-0.4}&\textcolor{red}{N}   & 1.04-1.31 &  >0.3.                                             &\textcolor{red}{0.30-0.40}.     & Y\\
  & & & & & &  & &  & & \\
8.0  & 450 & 7 & 0.88-1.07 & \textcolor{red}{0.00-0.40 or >9.3}  &\textcolor{red}{0.4-0.6 or >6.0}&       Y                    & 0.80-0.91         &  \textcolor{red}{<0.5 or >4.1}&0.4-0.8 or 2.8-7.2     & Y  \\
8.4  & 450 & 7 & 0.86-1.00 & \textcolor{red}{0.00-0.40 or >5.2}  &\textcolor{red}{0.4-0.6 or >4.6}&        Y                   & {\bf 0.76-0.91} & >0.0                                      & 0.4-6.0                    & Y \\
8.7  & 450 & 7 & 0.87-1.07 & \textcolor{red}{0.00-0.50 or >2.7}  &\textcolor{red}{0.4-0.6 or >3.4}&\textcolor{red}{N}& {\bf 0.75-0.91}  & >0.2                                      &0.5-5.0                     & Y \\
9.1  & 450 & 7 & 0.87-1.03 & >0.0                                               &\textcolor{red}{0.4-0.8 or >2.4}&\textcolor{red}{N}  & {\bf 0.77-0.92} & 0.3-9.5                                  &0.5-3.8                      & Y\\
  & & & & & &  & & & &\\
\hline\hline
\end{tabular}
\label{tablemod}
}
\end{center}
\end{table*}

\begin{figure*}
   \centering
     \includegraphics[width=9.5cm]{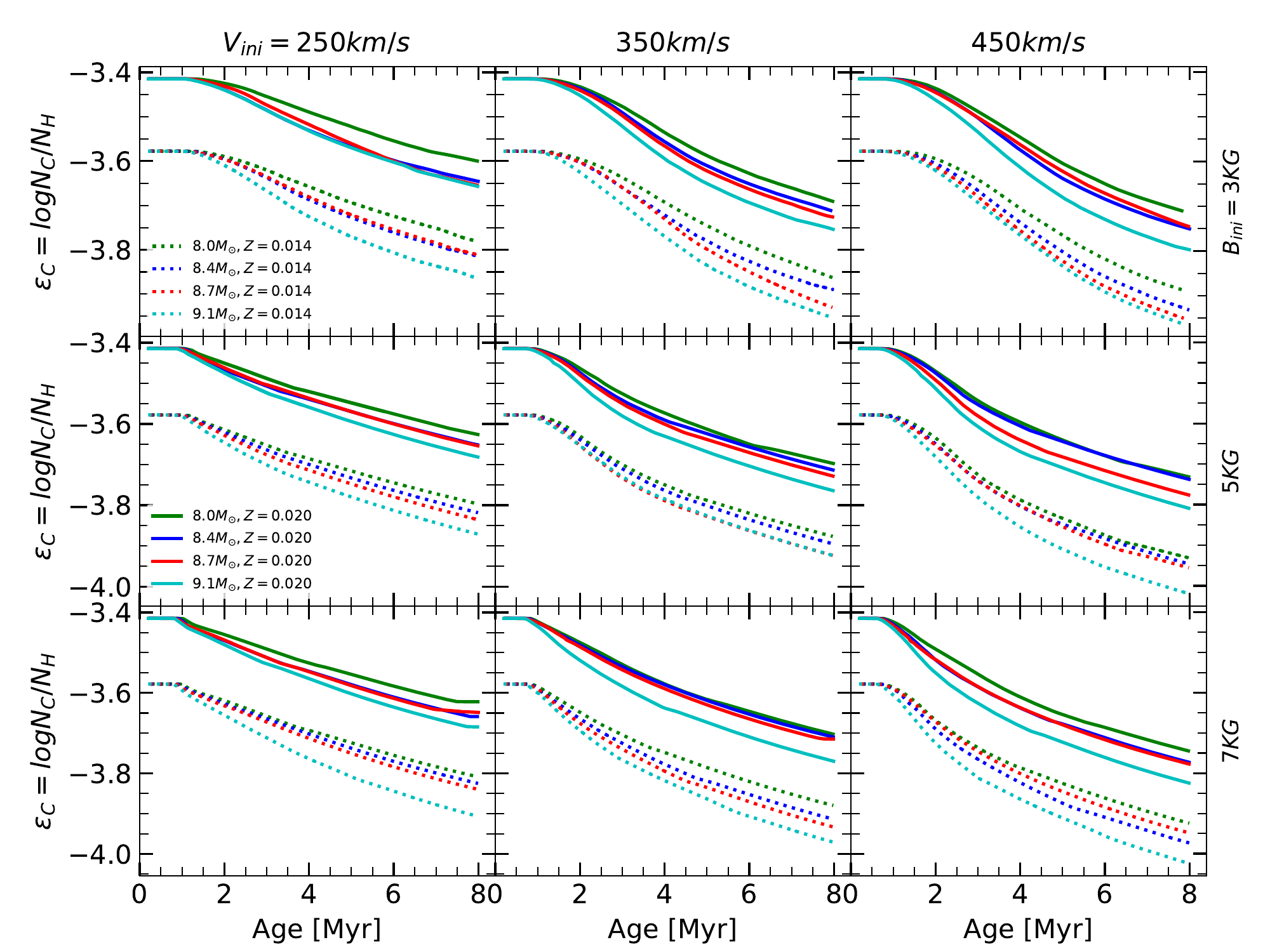}\includegraphics[width=9.5cm]{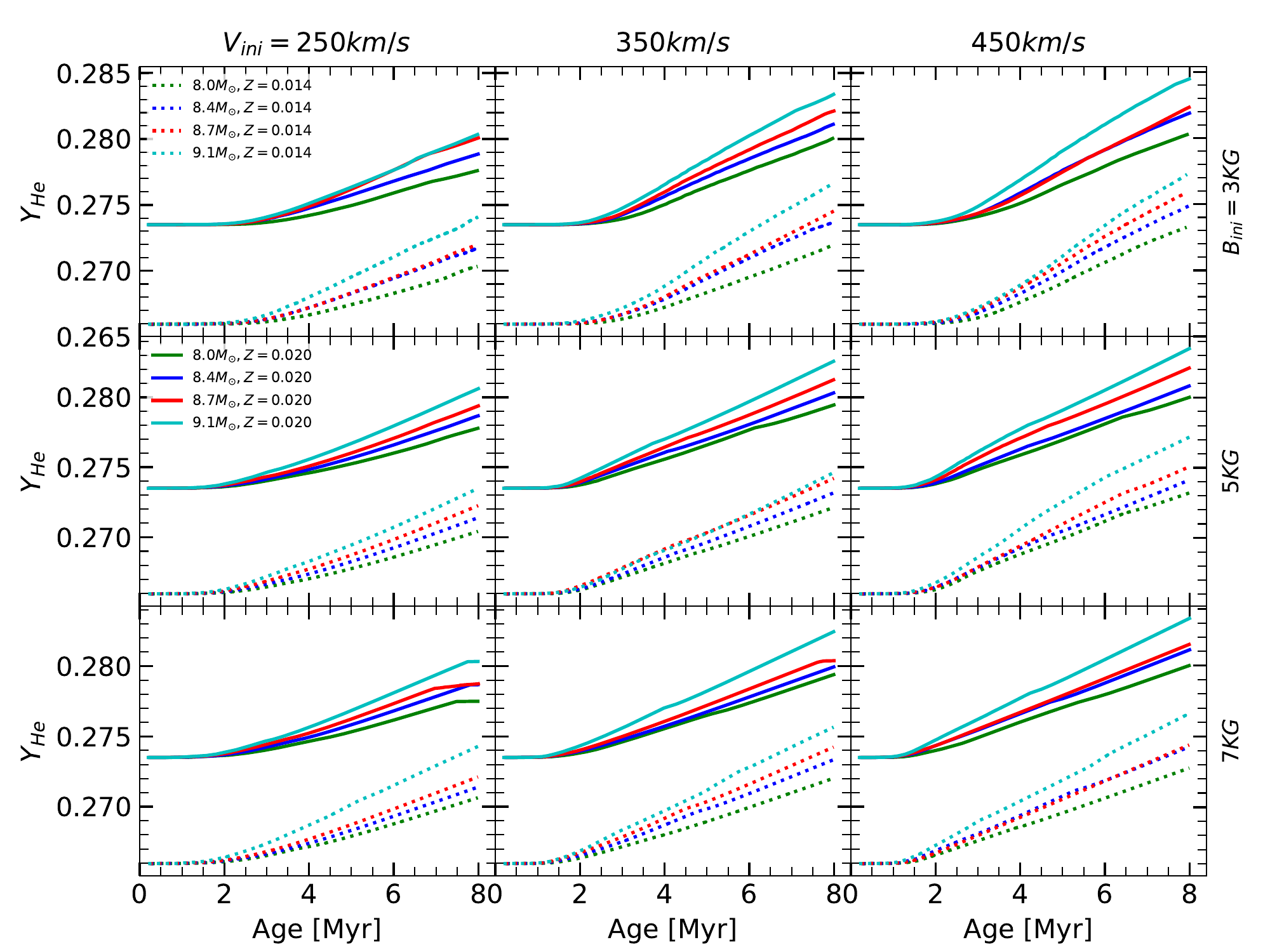}
      \caption{Evolution of the abundance ratios in number at the surface as a function of time. {\it Left panel:} the ratio of carbon to hydrogen.
      {\it Right panel:} the ratio of helium to hydrogen.
}
   \label{surfab}
   \end{figure*}

In Table~\ref{tableobs}, we have collected the data deduced from observations characterizing $\sigma$ Ori E (HD 37479).
The angular diameter, $\theta$=0.079$\pm$0.002 mas, has been obtained by \citet{GH1982} through the formula $\theta=2(F_{\rm earth}/F_{\rm theo.})^{1/2}$, where
$F_{\rm earth}$ is the integrated flux received on Earth, and $F_{\rm theo.}$ is a theoretical absolute flux determined using
Kurucz' model atmospheres.

Using the distance determined by Gaia (DR2, $439^{+20}_{-19}$~pc) and the angular diameter, one can determine the radius of $\sigma$ Ori E (3.73$\pm$0.26 R$_\odot$).
The effective temperature (22500$\pm$600 K) is deduced from the flux received on Earth and the angular diameter. This value is in agreement with the spectroscopic determination of the
effective temperature by \citet{Hunger1989}.
Note that a recent determination by \citet{Oksala2012}, using high resolution spectropolarimetry finds an effective temperature around
23000$\pm$3000 K\footnote{These authors however have not modeled in detail the spectrum and did not attempt to determine precisely the effective temperature.}, also not fundamentally different from the one obtained by \citet{GH1982}.
From the rotational period and the radius of the star, one can deduce the surface equatorial velocity (159$\pm$11 km s$^{-1}$).
Using the radius and the effective temperature, one can determine the luminosity of $\sigma$ Ori E quoted in Table~2.  This luminosity is in agreement 
with the spectroscopically determined one by \citet{Shultz2019a}, see Table~\ref{tableobs}. \citet{Shultz2019a} provides also
a value for the effective gravity also shown in Table~\ref{tableobs}.

We do not consider the surface magnetic field as well as the surface abundances as constraints of the present models.
As indicated in the previous section, the actual morphology of the surface magnetic field is more complex than the one used here to model
the wind magnetic braking. So the only quantity that we can hope to obtain is the value of an equivalent surface equatorial magnetic field
in a pure aligned dipolar morphology that can fit both the surface rotation and the spin-down timescale. The fitting value provides an
estimate of the field but can differ from the observed one.

$\sigma$ Ori E presents surface chemical inhomogeneities likely produced by microscopic diffusion that might occur in atmosphere stabilized by a strong magnetic field
\citep[see {\it e.g.}][]{Michaud1970}. In the entire class of Bp/Ap stars, these chemical inhomogeneities have been well documented for decades \citep{Babcock47, wolff68, Landstreet1978}.
Recently \citet{Panei2021} have explored the questions of chemical inhomogeneities at the surface of magnetic early B-stars.
Since these processes are not accounted for in the present models, we cannot use the observed surface abundances as constraints. On the other hand, we may  see whether rotational mixing is expected to have occurred in $\sigma$ Ori E. As we shall see, the age we obtain is too short for rotational
mixing as included in the present models to have a sensible effect.


\section{Physics of the models}

Stellar models are computed with GENEC, the Geneva stellar evolution code. The physical ingredients are the same as those of \citet{Ekstrom12} for what concerns
any non-magnetic effects. 
We use the Schwarzschild criterion for convection with a modest overshooting given by an extension of the radius of
0.1 $H_{\rm p}$, where $H_{\rm p}$ is the pressure scale height estimated at the Schwarzschild boundary. Rotational mixing is accounted
for according to the shellular theory by \citet{Zahn1992}. The diffusion coefficients and the physics of rotation are implemented as explained in \citet{Ekstrom12}. The mass-loss rate via stellar winds is used according to \citet{deJager1988}.
The present rotating models account for the various effects of a surface magnetic field on massive star evolution \citep[][]{Meynet11, Petit2017, Georgy2017, Zsolt2019, Zsolt2020}.

Similarly to previous GENEC implementations \citep[][]{Meynet11, Georgy2017, Zsolt2019}, we account for wind magnetic braking and mass-loss quenching. The wind magnetic braking is accounted for following the recipe given by \citet{Ud2009}. The present magnetic braking timescale of 1.3 Myr was found well in the range of values expected from the theory of wind magnetic braking given in \citet{Ud2009}
supporting thus the view that the observed braking might be due to that process. 

The rate of loss of spin angular momentum $\dot{J}_{\rm mb}$  due to magnetic braking is
expressed by
\begin{equation}
\dot{J}_{\rm mb}=\frac{2}{3}\dot{M}_{\rm wind}\Omega
R^{2}[0.29+(\eta_{\ast}+0.25)^{1/4}]^{2},
\end{equation}
where $\dot{M}_{\rm wind}$ is the mass-loss rate the star would have in absence of the magnetic field (this is $\dot{M}
_{B=0}$), $\Omega$ the surface angular velocity, $R$, the stellar radius,
$\eta_{\ast}=\frac{B_{\rm
eq}^2R^2}{\dot{M}\upsilon_\infty}$ the equatorial magnetic confinement parameter \citep{Ud-Doula2002} with $B_{\rm eq}$ the equatorial
magnetic field which is equal to half the polar field in case of a
dipolar magnetic field aligned with the rotational axis,
$\upsilon_\infty$ is the final wind velocity ({\it i.e.} the wind
velocity when there is no longer acceleration). The quantity $R[0.29+(\eta_{\ast}+0.25)^{1/4}]$
is the Alfv\'en radius, $R_{\rm A}$.

Since magnetic braking modifies the angular velocity of the stellar
surface, the Geneva code implements equation (1) as a boundary
condition of the internal angular momentum transport equation at
the stellar surface, and modifies the total angular momentum content
of the star.

We have implemented the effect of mass-loss quenching by the surface magnetic field
in the same way as Petit et al. (2017). We assumed that the magnetic field is constant in time (see the discussion in Sect.~5).
The escaping wind fraction $f_{B}$ (number inferior to 1) is taken as
\begin{equation}
f_{B}=\frac{\dot{M}_{\rm wind}}{\dot{M}_{B=0}}=1-\sqrt{1-\frac{R}{r_{c}}},
\end{equation}
where $r_{c}$ is the radius of the farthest closed loop of the magnetic field and is computed as a function of the Alfv\'en  radius
and the confinement parameter (for details, see \citep{Zsolt2017, Petit2017, Ud-Doula2002}. 
We did not account for the factor $1-\sqrt{1-0.5R/R_{\rm K}}$, where $R_{\rm K}$ is the Kepler corotation radius
defined by $R/W^{2/3}$, with $W=V/\sqrt{GM/R}$ \citep[see Eq.~22 in][]{udDoula2009}. The quantity $W$ is the ratio of the surface rotation to the Keplerian critical velocity {\it i.e.} the
velocity at which, keeping the stellar radius constant, the centrifugal acceleration balances the gravity at the equator.
Not accounting for this factor (as done here) overestimates the effect of the magnetic mass loss quenching. We note however that
this does not impact the angular momentum loss rate since this one depends on the mass loss rate in absence of any magnetic mass loss quenching.
The reason for this is that the mass retained in the magnetosphere slows down the star anyway and its effect is accounted for in the formula
for the angular momentum loss. Thus we suspect that actually the mass loss quenching has here a rather modest effect by modifying
the way the total mass of the star decreases. For the cases considered here the total mass removed remains anyway very modest.
As a numerical example, a 9 M$_\odot$ star has a mass loss rate between  10$^{-9}$ and 10$^{-10}$ M$_\odot$ per year during the MS phase.
Thus it loses a fraction of its total mass that is less than 0.3\% during the 30 Myr duration of the MS phase.
The magnetic mass loss quenching will still reduce that quantity. Typically considering for our 9 M$_\odot$ star, a surface magnetic field of 5 kG, an average surface
rotation during the MS phase of 200 km s$^{-1}$, we have that the Alven Radius $R_{\rm A}$ (about 30 times the stellar radius) is larger than the Keplerian Radius $R_{\rm K}$ 
(around 15 times the stellar radius). and we have thus here a centrifugal magnetosphere \citep{Petit2013}. In this case, rotation impacts significantly the dynamics of the magnetosphere \citep{T05}
and may lead to rotationally modulated variations of spectroscopic or photometric diagnostics  \citep[as e.g. Balmer lines, UV, X-rays, see eg.][]{Petit2013}.
When the magnetic mass loss quenching is accounted for such a star, the mass lost during the MS phase is only a few percents of the mass lost without that effect and amounts to 0.01\%
of the total mass of the star. Note the considering the data for $\sigma$ Ori E indicated in Table~1, assuming a mass around 9 M$_\odot$, one obtains that
the Alfven radius is around 50 stellar radii and the Kepler radius is around 30 stellar radii.

For single star models, we considered models of 8.0, 8.4, 8.7 and 9.1 M$_\odot$
with metallicities, $Z$=0.014 and 0.020, with initial rotations equal to 250, 350 and 450 km s$^{-1}$, and an equatorial
surface magnetic field of 3, 5 and 7 kG.


\section{Models for $\sigma$ Ori E}

Figures~\ref{veqtime} and \ref{pdotptime} present the evolution as a function of time of the surface equatorial velocity, of the stellar radius, of the braking timescale, as well as the evolutionary tracks in the HR diagram. The grey regions, in each panel, indicate the observed values of the different physical quantities.
In Table~\ref{tablemod}, each model is specified by its initial mass, rotation, surface
magnetic field (see columns 1 to 3), the age range where the model can fit the surface
velocity, the stellar radius and the braking timescale (see columns 3 to 4 for
the models with a metallicity equal to 0.014, and columns 7, 8 and 9 for the metallicity equal to 0.020). Columns 6 and 10 indicate whether the position of the star in the HR diagram can be fitted (Y) or not (N).
When for the age ranges, No Sol. is indicated, it means more precisely that there is
no solution in the age range between 0 and 10 Myr. We have used red colors to emphasize age
ranges that are not coincident with the one given by the requirement of fitting the surface
velocity. Boldfaced age ranges highlight the domains of ages allowing a simultaneous fit of
all the constraints indicated in Table~2.

\subsection{Theoretical predictions}

As shown in the left panel of Fig.~\ref{veqtime}, the surface equatorial velocity decreases
as a function of time. A consequence of the strong surface rotation decrease is that
very rapidly the star has a surface velocity that corresponds to a modest ratio of the critical velocity.
The critical velocity is the surface equatorial velocity the star should have in order for the gravity at the equator to be balanced by the
centrifugal force. As a numerical example, the critical surface velocity of stars\footnote
{
We assume here a Roche model where the critical velocity is given by 
$\sqrt{2G M \over 3 R}$.
} 
with an initial mass between 8 and 9.1 M$_\odot$
and a radius compatible with the one indicated in Table~2 for $\sigma$ Ori E is between 506 and 578 km s$^{-1}$. Models that
satisfy the observed constraints have surface velocities between 150-170 km s$^{-1}$.  This represents a surface velocity
equal to 26-34\% of the critical values. This is a too low ratio for rotation to have  any significant effect on the shape of the outer layers.

From the left panel of Fig.~\ref{veqtime}, we see that for a given initial rotation, the results do not show a great sensitivity
on the changes of the initial mass in the narrow mass interval between 8.0 and 9.1 M$_\odot$. 
Increasing
the surface magnetic field implies, as expected, a stronger braking. A given surface velocity is
thus reached at an earlier time. 
The models at $Z$=0.014 shows a higher surface rotation at a given
age than those at $Z$=0.020. When the metallicity decreases, stars are more compact and have weaker stellar winds, this decreases, at a given surface rotation rate, the loss of angular momentum
and thus increases the braking timescale.
Let us recall that a scaling of the wind with metallicity of the form $\dot{M}_{\rm wind}\propto (\frac{Z}{Z_{\odot}})^{0.5}$ has been used here \citep[as in the Geneva grids of stellar models][]{Ekstrom12, Georgy13a}. 

The evolution of the stellar radii is shown in the upper panel of Fig.~\ref{veqtime}.
Each curve shows two parts: a first short phase where the radius decreases and a second long phase during which it increases.
The first phase is due to the very efficient slowing down of the star by the magnetic wind braking effect.
To understand this, it is important to remind that rotation deforms the star making it oblate. Thus the radius plotted in this panel is actually an average radius, defined as
the radius of a spherical star that would have the same volume as the rotationally deformed star.
Since the polar radius is not significantly changed
by rotation \citep[see Fig.~2 in][]{Ekstrom2008}, and since the equatorial radius increases when rotation increases, the volume of a star increases with rotation \citep{Zsolt2020}. When the star is braked down, its volume decreases. This explains the first short decreasing phase. At a given time, the braking timescale becomes long enough for the secular evolution of the star to become the main agent driving the evolution of the radius. As is well known, during the MS phase the radius increases and this is what we see in that second phase. The first phase last longer for the models starting with a high initial rotation. It becomes shorter for the models with a high surface magnetic field. As is well known, models at $Z$=0.014 have smaller radii at a given age than models at $Z$=0.020 (the difference of stellar radius is approximately 0.2 $R_{\odot}$ between the two models with different metallicities).

The evolution of the braking timescale is shown in the left panel of Fig.~\ref{pdotptime}. We can identify the two phases coming from the evolution of the radius just described above. In the first phase, the
braking timescale increases rapidly due to the rapid slowing down of the star. It reaches a maximum then decreases. The decrease is mainly due to the fact that the mass loss rate and the stellar radius increase during the evolution along the MS phase. Models with $Z$=0.014 show longer braking timescales than those with $Z$=0.020. At lower metallicity the stars are more compact and the stellar winds are weaker thus the angular momentum loss rate, at a given surface velocity, are weaker.
We see that, for each model, a given braking timescale, in a fixed range different for each model, may occur at two different ages of the star. 
One at the very early time and one at a time when the age of the star is significantly above the braking timescale. This is due to the evolution described above.

In the lower panel of Fig.~\ref{pdotptime}, we also see that lowering the metallicity shift the tracks to hotter parts of the HR diagram. We see  the consequences on the HR diagram of the
short first phase during which, due to braking the average radius decreases. This corresponds to the phase, at the very beginning of each track, that is nearly horizontal.
The luminosity keeping constant, the effective temperature increases. Note that the effective temperature defined here is also an average effective temperature. Indeed, due to the von Zeipel theorem \citep{Zeip1924}, the effective temperature varies as a function of the latitude when the star is rotating, the poles being hotter than the equatorial regions. The effective temperature plotted in the
figure is obtained as $L/(4\pi R^2)$, where $R$ is the average radius.

\subsection{Comparisons with $\sigma$ Ori E}

Using Figs~\ref{veqtime} and \ref{pdotptime} (only the top panel), we can derive the range of ages of the models when they are inside the grey zone, {\it i.e.} fit the corresponding property of $\sigma$ Ori E deduced from the observations. The bottom panel of Fig.~ \ref{pdotptime} does not directly allow to deduce the age range (although some constant age lines are shown) but allows us to see whether
the beginning of the tracks goes through the observed position of $\sigma$ Ori E in the HR diagram.

The constraint coming from the surface velocity points towards a young star (see columns 4 and 8 of Table 3) with an age between 0.4 and 2.5 Myr for all masses, initial rotations, metallicities or surface magnetic fields considered here.
Among the other three constraints, the one on the magnetic braking timescale (see columns 6 and 10) is the most efficient in eliminating many models.

For the metallicity $Z$=0.014, actually no models is found having a $P/\dot{P}$
compatible with the one observed in the age range given by the observed surface equatorial velocity (although sometimes the miss is due to a rather small incompatibility).
The constrain of the radius at $Z$=0.014 points to older ages than those needed to account for the surface velocity except in the case of the 8.7-9.1 M$_\odot$ stellar models that are compatible with very young ages. The position in the HR diagram at $Z$=0.014 favors masses around 8.0-8.7 M$_\odot$.

For the metallicity $Z$=0.020,  only models with a surface equatorial magnetic field around 7 kG provide a simultaneous fit to the contraints indicated in Table~2. Depending on the initial rotation, the age
is either between 0.4-0.5 (250 km s$^{-1}$),  0.6-0.7 (350 km s$^{-1}$) or 0.75-0.9 (450 km s$^{-1}$). As already mentioned above, for a given surface magnetic field, starting from a larger rotation allows the star to reach the observed surface velocity at a later time. At $Z$=0.020, except for some models starting with 250 km s$^{-1}$, the position in the HRD can always be fitted.

The finding of consistent solutions is thus easier at $Z$=0.020 than at $Z$=0.014. In Table~3, we have highlighted in boldface the range of ages where all the five constraints shown in Table~2 can be fitted.
As mentioned above, only models at $Z$=0.020 are found. Actually we cannot say that models with a metallicity $Z$=0.014 reproducing the observed properties of $\sigma$ Ori E do not exist, because
we did not explore changes in the mass loss rates, convective core size, different angular momentum transport for instance. But it would bring us too far to explore all these possibilities, especially because,
as discussed below, a metallicity of $Z$=0.020 for $\sigma$ Ori E is not unreasonable at the moment. 

Taken at face, the above results show that the models that best match the observed properties of $\sigma$ Ori E have the following properties: the initial mass
is between 8.4 and 9.1 M$_\odot$, the metallicity is around $Z$=0.020, the age is between 0.4 and 0.9 Myr, and finally the magnetic braking is similar to the one due to a dipolar aligned magnetic field with
an equatorial value of $\sim$7 kG. The high surface rotation and magnetic field of $\sigma$ Ori E is possible because it is a very young single star.

We could have used as an additional constraint the surface gravity that is 4.2$\pm$0.2 (see Table~1) according to \citet{Shultz2019a}, but it would not help constraining more the model.
Indeed for all the four initial masses considered here, between 8.0 and 9.1 M$_\odot$, the minimum and maximum radii allowed by the measured surface gravity define a region that
overlaps and extends beyond the shaded region shown in the bottom panel of Fig.~\ref{veqtime}. Thus any solution fitting the radius will fit the constraint of the surface gravity.


Figure~\ref{surfab} shows the evolution of the surface abundances of carbon and helium as a function of time. We see that the changes in helium abundances is
very small with respect to those of carbon (the abundance of carbon is normalized to that of hydrogen, while the abundance of helium is not. However hydrogen does not change much and most of the variation shown for carbon is due to the change of carbon only). For carbon, models predict a decrease by a factor 2 (0.3 dex)
already after 8 Myr and for the models with an initial rotation larger than $\sim$350 km s$^{-1}$. 
Helium needs much more time than carbon to show changes at the surface.
This effect is due to the following facts: first let us recall that the diffusive velocity for a given element $i$ scales as ${1 \over X_i} {\Delta X_i \over \Delta r}$, where $X_i$ is the mass fraction of the element
and $\Delta X_i /\Delta r$ is the gradient of the abundance of this element. At the beginning there is no gradient, Zero Age Main-Sequence models being chemically homogeneous. In absence of any mixing in the radiative zones, the gradients between the core and the envelope then evolve under the actions of both nuclear burning and convection. Helium increases at the center
and carbon decreases due to the action of the CN cycle. The CN cycle is very rapid, much more rapid than the synthesis of helium. After one million year,  in a 9 M$_\odot$, the difference
between the mass fraction of helium in the core and that in the envelope is 0.11 (in the core one has a mass fraction of 0.277 and in the envelope of 0.266). For the carbon we have in the core
a mass fraction of 0.00003 and in the envelope the value is 0.0023. A rough estimate of the ratio of the diffusive velocity over the same value of $\Delta r$ will therefore be $V_{\rm He}/V_{\rm C} \approx 0.011/0.271 \times 0.0012/0.0023 \sim 0,02$. Thus the diffusive velocity of helium is about only 2\% the diffusive velocity of carbon, hence the surface will show signs of the CN cycle before any
significant change of helium.

The value quoted by \citet{Oksala2012, Oksala2015} (see Table~1 and Sect.~5.2) for the abundance of carbon (between
-~4.0 and -~5.0 dex, see table 1) cannot be reached at any time by the present models (actually the most massive models at $Z$=0.014 can just reach value just below -4 dex after 8 Myr).
It would be very interesting to have data for the nitrogen abundances to see whether this low carbon might be associated with an increase in nitrogen abundances. This would indicate the presence at the surface of material having been processed by the CN cycle, that very rapidly reaches equilibrium at the centre.

The observed helium surface abundance available at the moment, ($\epsilon_{\rm He}$ between -1.1 dex and -0.6 dex) span a very large domain.
$\sigma$ Ori E presents a non homogeneous surface composition due to processes as microscopic diffusion in regions of the star where a strong magnetic field allow to stabilize the atmosphere. Atomic diffusion like in Ap/Bp stars would mostly impact heavier elements, like Si and Fe-group, producing an anomaly with an overabundance in some and an under-abundance in others.
We have not accounted for such processes in our models thus we can not make comparisons with the observed surface abundances. On the other hand, Fig.~\ref{surfab} shows that,
for the very small ages obtained for $\sigma$ Ori E using the present models (below 1 Myr), no changes of the surface composition due to rotational mixing (as accounted for in the present models) is expected.

\section{Discussion}

\subsection{The age of $\sigma$ Ori E and the age of the $\sigma$~Orionis open cluster}

\begin{figure}
   \centering
     \includegraphics[width=6.0cm]{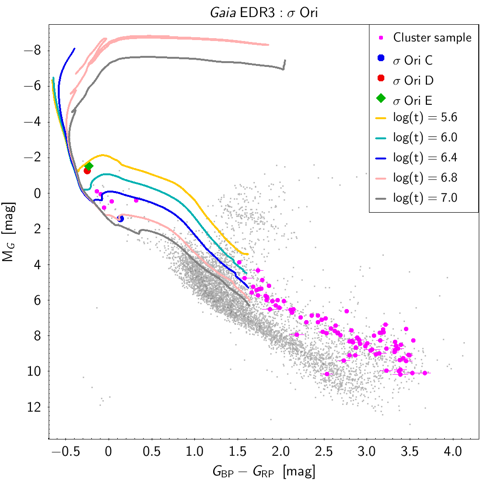}
      \caption{Observed positions in the \textit{Gaia} observational HR diagram of potential (see text) members of the $\sigma$ Orionis cluster (black filled circles), with \textit{Gaia} sources in the cluster field of view with parallaxes better 10\% shown in small gray points in the background. The uncertainties due to both photometry and parallax uncertainties are shown for all cluster members, being for the majority of them smaller than the size of the points. Superposed are shown isochrones for a metallicity Z=0.014 from \citet{Haemerle2019} for different values of the logarithm of the age (in years). The positions of 
$\sigma$~Ori C, D and E are highlighted in the diagram as shown in the inset. 
}
   \label{figage}
   \end{figure}

Since the star $\sigma$ Ori E belongs to the $\sigma$ Ori cluster, we can wonder whether estimating the age of the cluster through isochrone fitting
gives values that are compatible with the age deduced above for $\sigma$ Ori E  based on fitting the surface rotation, the radius, the breaking timescale and the position in the HR diagram. Let us remind that
\citet{Sherry2008, Caba2007} have obtained using this technics ages between 2-3 Myr, thus larger than the ages inferred for $\sigma$ Ori E in the present work.
However, as is well known, age determination is a very model dependent process.

We present in Fig..~\ref{figage}, the positions in the observational HR diagram of a sample of member candidates to the $\sigma$~Ori cluster using \textit{Gaia} EDR3 photometry and parallax.
The member candidates are selected from an initial sample of sources within a cercle of 600~arcsec radius around the center of the cluster center, with the conditions $2 < \varpi \mathrm{\;[mas]} < 3$ on the parallax, $0.2 < \mathrm{PM}_\mathrm{RA} \mathrm{\;[mas/yr]}< 2.5$ on the right-ascension proper motion and $-2.2 < \mathrm{PM}_\mathrm{DEC} \mathrm{\;[mas/yr]}< 0.5$ on the declination proper motion.
These selection criteria favor purity of cluster membership rather than completeness.
The candidates have further been restricted to those brighter than 19~mag in $\textit{G}$ and having uncertainties in $\textit{G}_\mathrm{BP}-\textit{G}_\mathrm{RP}$ less than 0.1~mag.
The $\textit{G}_\mathrm{BP}-\textit{G}_\mathrm{RP}$ uncertainties and the ones of the absolute \textit{G} magnitude are shown in Fig~\ref{figage}. with horizontal and vertical bars, respectively, the latter  uncertainty including both photometric and parallax uncertainties.
These uncertainties are smaller than the size of their data points in the figure for the great majority of the member candidates.
It must be noted that past studies of this cluster conclude on very little reddening of its members \citep{Bejar01,Oliveira02}.
We therefore did not apply any reddening correction.

Superposed are shown isochrones for different ages between 0.4 Myr ($\log (t)=5.6$) and 10 Myr ($\log (t)=7.0$) accounting for the pre-Main-Sequence phase
computed with accretion and no rotation by \citet{Haemerle2019} and a metallicity $Z$=0.014\footnote{At the moment no such isochrones are available for a metallicity $Z$=0.020. A value of $Z$=0.020 would shift the isochrones slightly to the right in Fig.~\ref{figage}.}. The isochrone passing through the observed position of $\sigma$ Ori E is either an isochrone corresponding to a very young age with a log age between 5.6 and 6.0 or an isochrone with a log age larger than 7 (not shown here).  However, a young age is likely the most reasonable solution in case most of the stars below a magnitude G equal to 4.0 are indeed pre-MS stars. Thus it appears that an age of 1 Myr for $\sigma$ Ori E  as deduced above, based on fitting the observed properties of $\sigma$ Ori E,
does not appear in contradiction with the age determined by isochrone fitting. We note however that this aspect would need a more careful analysis and also a trial with similar isochrones but computed with a metallicity $Z$=0.020.

\subsection{The metallicity of $\sigma$ Ori E}

A solution is obtained only if the metallicity is around $Z$=0.020. A value of $0.014$ makes the finding of a solution more difficult if not impossible.
Since $\sigma$ Ori E presents surface chemical inhomogeneities, direct determinations
of its metallicity may be problematic. \citet{Oksala2015} in their Table~1 indicate values for $\epsilon_{\rm Fe}$ (=$\log (N_{\rm Fe}/N_{\rm H})$) between -5.7 and -4.0 for $\sigma$ Ori E.
For comparison, the solar values in the same units as in \citet{Oksala2015} given by \citet{Scott2015} is -4.53 . 
The large domain given by \citet{Oksala2015} does not allow to conclude whether $\sigma$ Ori E is metal deficient or, on the contrary, slightly metal-rich compared to the Sun. 
Also, as already mentioned above, the Oksala's abundances may not reflect the bulk iron abundance in the star and 
result of some diffusion processes unevenly affecting its surface composition.
To obtain
an idea of the actual metallicity, it is better to rely on measurements of normal stars belonging to the same association.  According to \citet{Cunha1998}, the star HD 294297 belongs to the Ori OB 1b Association
where the $\sigma$ Orionis cluster lies. They obtain an iron abundance ($\epsilon_{\rm Fe}$) of $-4.68\pm0.14$, thus below the solar abundance. If we assume that $Z/Z_\odot=10^{\epsilon_{\rm Fe}-\epsilon_{\rm Fe_{\odot}}}$ (i.e. assuming a solar scaled distribution of the heavy elements), then it would mean that $Z$ would be equal to 0.71$Z_\odot$, thus 0.010 with $Z_\odot$=0.014. Taken at face, if this metallicity is also the one corresponding to $\sigma$ Ori E then our $Z$=0.020 model would not be an acceptable solution.
In that case, either other parameters of the models should be changed to check whether a solution can be found at a metallicity lower than 0.020, or a more complex scenario involving
multiple stars, like, for instance, a merging of two stars has to be invoked. However at the moment
it is difficult to discard the $Z$=0.020 single star evolution scenario on the basis of this metallicity measurement. If we look at Table 2 of \citet{Cunha1998}, and compare the iron abundances for stars belonging to one association (here Ori OB 1c), it goes from -4.8 (0.54 $Z_\odot$= 0.008 applying the same rule as above)  up to -4.41 (1.32 $Z_\odot$= 0.018), which shows variations by a factor of 2.5! Moreover, \citet{Cunha1998}
finds that the abundances of oxygen shows still greater diversity of abundances than iron. They suggest that some regions where the observed stars have formed have been enriched by a nearby supernova.
This would have an impact on the mass fraction of heavy elements $Z$. Thus at the moment it is still difficult to reach a very strong conclusion until more data will be collected on the Ori OB 1b Association.

In this work we have not considered different initial helium mass fractions at a given metallicity. Decreasing helium at a given metallicity shifts the position of the evolutionary tracks in the HR diagram to the red
may be allowing to find a solution for a metallicity equal to Z=0.014.

\subsection{The magnitude of the surface magnetic field}

The observed polar surface magnetic field of $\sigma$ Ori E has been obtained by \citet{Oksala2012, Oksala2015}. According to \citet{Oksala2015}, there is a dipolar component that is misaligned with respect to the rotation axis. The polar strength of this magnetic field is between 7.3 and 7.8 kG, with an obliquity between 47$^\circ$ - 59$^\circ$ but there is also a smaller non-axisymmetric quadrupole component with strength between 3-5 kG. Ideally of course one would need to produce a detailed modeling of the braking law resulting from the actual magnetic field topology.
At the moment, based on the present simulations, we can only say that using an aligned dipolar configuration, we need a magnetic field polar strength that is around 14 kG, so larger than the observed one. Would a more realistic topology need a smaller polar field more in line with the observed one? Likely not, since both the misalignment and quadrupole components would be expected to decrease the braking efficiency overall compared to a pure aligned dipole. So this question at the moment remains open. Note that in this work we have not explored the possibility that the magnetic field flux may decay with time  \citep[discussions of  possible underestimates of the braking efficiency and of the effect of magnetic flux decrease is discussed by][in their study of the B-type star tau Sco]{Zsolt2021}. However the constraints given by the surface velocity and the isochrone fitting favors anyway a young age and thus a rather limited impact of this effect.

\subsection{The timescale for diffusion}

The star $\sigma$ Ori E shows sign of the action of microscopic diffusion. One can wonder whether the small age obtained gives enough time 
for such a process to impact the surface composition.
According to Fig. 8.1 of \citet{Michaud2015}, the diffusion timescales in a stable atmosphere of a 2.5 M$_\odot$ star
are between 10$^3$ and 10$^4$ years depending on the element considered
(the diffusion timescale is the time for a given chemical element to diffuse over a pressure scale height). These timescales are much  shorter than the age derived here for $\sigma$ Ori E. However,
these estimates are based on a non-magnetic atmosphere and are valid for stars with an effective temperature less than 16 000 K. Above these temperatures, stellar winds
are expected to prevent any stratification due to diffusion. In case of $\sigma$ Ori E, the surface chemical inhomogeneities may  be associated with regions stabilized by the strong magnetic field. Interestingly,
the presence of a magnetic field does not change much the timescales  indicated above \citep[Georges Alecian, private communication and see also Fig. 1 in][]{Stift2016}. Thus, from this discussion,
we conclude that diffusion has likely enough time to operate, at least in some stabilized regions, in timescales less than 1 Myr.

\section{The $\sigma$ Ori E analogs}

\begin{figure}
\centering
\includegraphics[width=9.8cm]{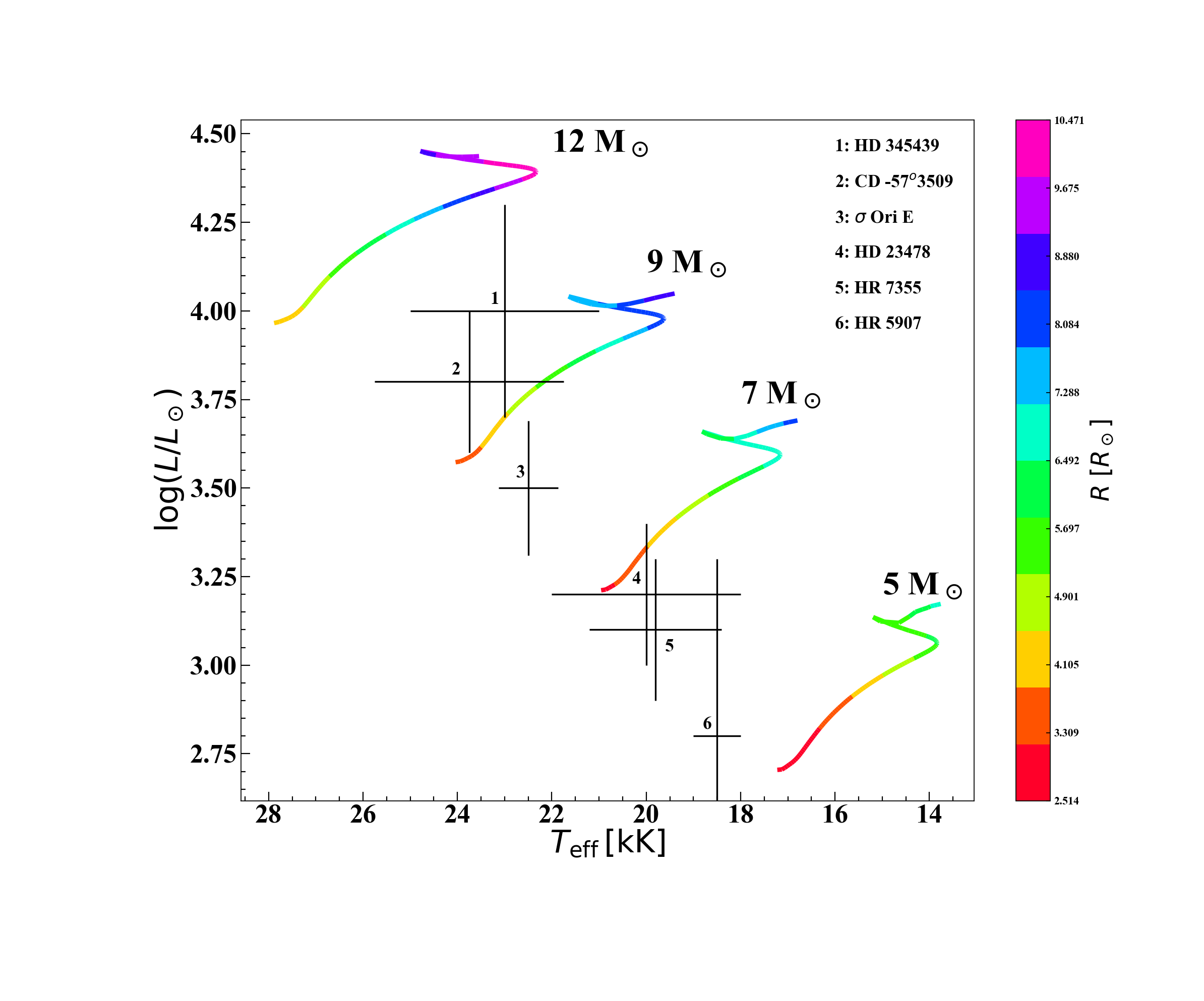}
      \caption{Positions of $\sigma$ Ori E  and analogous stars in the HR diagram. The data for $\sigma$ Ori E are those indicated in Table 2. For the other
      stars, the data have been taken as given in Table  1 of \citet{Shultz2020}, except for the star CD -57$^{o}$3509 taken from \citet{Przy2016}.
      Main-Sequence rotating tracks from \citet{Ekstrom12} for the 5, 7, 9 and 12 M$_\odot$ are indicated.}
         \label{HRDanlaogs}
  \end{figure}

\begin{table}
\scriptsize{
\caption{Characteristics of some $\sigma$ Ori E analogs ordered by decreasing $\upsilon\sin i$.}
\label{sigma}
\begin{center}
\begin{tabular}{ccccc}
\hline
Name                & type                 & $\upsilon\sin i$ & B$_p$    &  References   \\
                         & type                 & [km s$^{-1}$]    & [kG]        &   \\
\hline
HR 7355          & B2Vnp                   &   300$\pm$15   & 11-12    &  [1]           \\
HR 5907          & He strong CP   &   290$\pm$10   & 10-16    &  [2]     \\
HD 345439       & He-rich   B2V   &  270$\pm$20    & -  & [3]             \\
$\sigma$ Ori E  & B2Vpe              &  160                  & 7.3 - 7.8 & [4]            \\
HD   23478       & He-norm B3IV  &  125$\pm$20    & $\ge 9.5$  & [5] [6]           \\
CPD -57$^{o}3509$ & B2IV         & 70-250              &  3.3 &  [7]  \\
\hline
\multicolumn{5}{l}{[1] \citet{Rivi2008,Rivi2013}; [2] \citet{Grunhut2012}} \\
\multicolumn{5}{l}{[3] \citet{Eikenberry2014}; [4] \citet{Oksala2015}} \\
\multicolumn{5}{l}{[5] \citet{Eikenberry2014} [6] \citet{Sikora2015}; [7] \citet{Przy2016}}  \\
\end{tabular}
\end{center}
}
\end{table}

In this section, we briefly discuss a few cases of $\sigma$ Ori E analogs, {\it i.e.} of B-type stars presenting both a high surface rotation and a strong surface magnetic field. 
The positions of these stars in the HR diagram are shown in Fig.~\ref{HRDanlaogs}. 
In Table~4,  some properties of such stars are listed together with those of $\sigma$ Ori E. From the mean positions of these stars in Fig.~\ref{HRDanlaogs}  (not accounting for the error bars), we can deduce the mass and radius of each star as would be given by the stellar models and hence their critical velocity. Then from the $\upsilon\sin i$ indicated we obtain the minimum ratios of the actual equatorial to the critical velocity. 
We obtain values  between about 25\% (HD 23478)
and 70\% (HR 5907). According to \citet{Ekstrom2008}, a 70\% ratio makes the equatorial radius to be about 20\% larger than the polar one, implying thus
some significant deformation.

The fast spinning stars with a strong surface magnetic field can be classified into two categories: 
those for which there are evidences that they are so young that fast rotation together with a strong surface magnetic field is not actually challenging single star models (like $\sigma$ Ori E) and those for which evidences exist for a sufficiently large age or advanced evolutionary stage that present single star models cannot account for their properties. This second category requires either
that the wind magnetic braking law does not apply or that an interaction with a companion spin-up or has (in case of a merging event in a recent past)  spin-up the star.
Looking at Fig.~\ref{HRDanlaogs}, most of these stars have positions in the HRD compatible with a young age, 
except HD 345439 where the uncertainties do not allow for drawing a firm conclusion regarding its evolutionary status. 
In the discussion below, let us keep in mind that the age determinations are obtained from various different modeling assumptions, making this quantity highly uncertain.

HR 7355 (HD 182180) is a helium-strong chemically peculiar star with Balmer emission lines \citep{Rivi2008, Rivi2013, Ok2010}. It shows He-strong absorption and polar field strength between 11 and 12 kG. Its rotation is exceptionally fast with a spin period of 0.52 days \citep{Rivi2008, Miku2010}, implying a surface velocity $\upsilon_{\rm eq}$ = 310$\pm$5 km s$^{-1}$ \citep{Rivi2013}.  
\citet{Miku2010} estimate the age to be between 15 and 25 Myr and a mass of 6.3$\pm$0.3 M$_\odot$ using isochrones from \citet{Mar2008}\footnote{\citet{Ok2010} obtained a gravity log $g$ equal to 3.95 that would support a younger age for this star.}.  
\citet{Miku2010} found a characteristic braking timescale of about 0.4 Myr, under the assumption that the star is a solid-body rotator. This short timescale would imply that the star is rapidly undergoing very efficient magnetic braking, suggesting a young age\footnote{\citet{Miku2010} have indeed possibly found evidence for a lengthening of the rotational period with
$\dot{P}/P=2.4(8)$ 10$^{-6}$ yr$^{-1}$ during the last 20 years, {\it i.e.} 108 ms per year.}. Thus this would be a challenging case for single star evolution!
However, a study by \citet{Rivi2013} showed that this case might not be so challenging after all when proper account has been made of the gravity darkening effects \citep[see {\it e.g.}][]{Zeip1924, Lara2011} that are certainly important for such a fast rotating star \citep[see Fig.~2 in][]{Georgy2014}. \citet{Rivi2013} redetermined the properties of this star accounting for these effects and found a significantly lower mean effective temperature than previously found. From that lower
effective temperature, they deduce weaker winds and thus longer magnetic wind braking timescale. These authors conclude in contrast with previous works that the age and rapid rotation are not inconsistent with the presence of a fossil magnetic field. This case illustrates the fact that at high rotation, one cannot neglect the effects of the deformation of the star.

According to \citet{Grunhut2012},
HR 5907 (or HD 142184 or V1040 Sco)  has a spin period of 0.51 day and an equatorial surface velocity of 340 km s$^{-1}$ \citep{Fremat2005}; the star has an effective temperature of 17000$\pm$1000 K, a projected rotational velocity of 290$\pm$10 km s$^{-1}$, an equatorial radius of
3.1$\pm$0.2 R$_\odot$ and a stellar mass of 5.5$\pm$0.5 M$_\odot$, a surface dipole field strength between $\sim$10.4 and 15.7 kG. (V=5.4, B2.5V). It is located in the Upper Scorpius OB association at a distance of $\sim$ 145 pc\footnote{
The \Gaia parallaxes confirm this distance for HR 5907, with a parallax of $7.08 \pm 0.14$~mas from DR2 ($141 \pm 3$~pc), and $6.990 \pm 0.074$~mas from EDR3 ($143.06 \pm 0.14$~pc).
}
\citep{Her2005}. \citet{Grunhut2012} also computed the spin-down timescale and found a value of about 8 Myr which is longer than the estimated age of HR 5907 
by \citet{Her2005}, but shorter than the age estimate of 10 Myr by \citet{Feid2016}\footnote{Actually these authors give age estimates for the Upper Scorpius association to which HR 5907 belongs.}
So depending on the age estimate, this star may or may not be a challenging case for single star models.

Two stars HD 345439 and HD 23478 have been discovered serendipitously in the course of the APOGEE (Apache Point Observatory Galactic Evolution Experiment) survey \citep{Eikenberry2014}. They detected the characteristic double-horned profile of emission lines in the Brackett series. This feature comes from material trapped in a magnetosphere rotating rigidly with the star \citep[model of Rigidly Rotating Magnestosphere, RRM, see][]{Townsend2005, Townsend2010}.
For HD 345439 these authors deduced from the line profile of He I absorption feature $\upsilon\sin i = 270\pm20$ km s$^{-1}$.
They could not measure from their data the magnetic field, however the presence of a strong magnetic field is made evident through the presence of this RRM.
\citet{Hubrig2015}, using four subsequent low-resolution FORS 2 spectropolarimetric observation of that star,  did not detect any magnetic field at a significance level of 3$\sigma$. However this null result may be due to a rapidly
varying magnetic field. Analysis of the four individual spectropolarimetric observations of that star is compatible with a variation of about 1 kG of the longitudinal magnetic field in a period of 88 minutes.
Its spectral type indicates that it might be a star with a mass similar to $\sigma$ Ori E. It is difficult to say whether this star poses a challenge in the sense that little is known about its age.





HD 23478 is as HD 345439 a fast rotating star of the $\sigma$ Ori E type that shows the  presence of a rigidly rotating magnetosphere (RRM) \citep[for a detailed study see][]{Sikora2015}. However it is a ``He-normal'' B3IV  star \citep{Eikenberry2014}.
A  $\upsilon\sin i = 125\pm20$ km s$^{-1}$
is obtained from the line profile of He I absorption feature\footnote{A previous work obtained a photometric period of 1.0499 days which is slightly faster than the 1.19 days rotation period of $\sigma$ Ori E \citep{Jery1993}.}. Its sky position and proper motion are compatible with a belonging of that star to the IC 348 young open cluster \citep[proper motions of stars in this cluster have been studied by][]{Scholz1999}. If indeed it is a member of that cluster, then the age of that star would be between 1.3-3 Myr according to \citet{Herbig1998} and between 5-6 Myr according to \citet{Bell2013}. \citet{Hubrig2015} performing low-resolution FORS 2 spectropolarimetric observations of that star discovered a rather strong longitudinal magnetic field of up to 1.3-1.5 kG (this is the magnetic field along the line of sight). This star might be an interesting candidate to study in a similar way as $\sigma$ Ori E. It would be interesting to have data on the braking timescale.

CPD -57$^{o}3509$ (B2IV) is a member of the Galactic open cluster NGC 3293 that has an age of about 8 Myr \citep{Baume2003}. \citet{Przy2016} detected a surface averaged longitudinal magnetic field with a maximum amplitude of about 1 kG. They deduce this star has a bipolar magnetic field with a strength larger than 3.3 kG (assuming a dipolar configuration). They also observe large and fast amplitude variations (within about 1 day) of the longitudinal magnetic field. They interpret this fact as reflecting a very fast rotation (although the projected rotational velocity is small, around 35 km s$^{-1}$). The star shows no sign of a RRM. They obtain an effective temperature of 23750$\pm$250 K and a log $g$ of 4.05$\pm$0.10. Using the Geneva track \citep{Ekstrom2012}, they deduce a mass of 9.7$\pm$0.3 M$_\odot$, a radius of 5.0$\pm$0.9 R$_\odot$ a Log $L/L_\odot$ = 3.85$\pm$0.13 and an age of 13.8+2.4-3.3 Myr. The characteristics deduced from the models by \citet{Brott2011} are respectively 9.2$\pm$0.4 M$_\odot$, 4.4+0.7-0.5 R$_\odot$, log $L/L_\odot$=3.76$\pm0.12$ and an age of 13.0+1.7-4.0 Myr. They provide some range for the equatorial velocity between about 70 and 250 km s$^{-1}$. This star might also be an interesting case for checking the process of wind magnetic braking. An important additional piece of information would be to determine $\dot{P}$. However, given the high inclination of the star, it seems unlikely we will get $\dot{P}$ any soon.

\section{Conclusions and future perspectives}

We have revisited here the case of $\sigma$ Ori E taking benefit from more accurate distance determinations provided by Gaia and
a new series of computations that account for both the wind magnetic braking and the magnetic mass loss quenching. 

We obtain that $\sigma$ Ori E is a very young star (age less than 1 Myr) as was obtained by \citet{Townsend2010}.
The mass of $\sigma$ Ori E is between 8 and 9 M$_\odot$ and its metallicity $Z$ (mass fraction of heavy elements) around 0.020.
The braking law can well reproduce the observed slowing down in the frame of an aligned dipolar magnetic field topology. We obtain in that case
that the polar magnetic field needed is around 14 kG thus two times larger than the observed one. We have not succeeded to resolve this
discrepancy. We just noted that the actual magnetic field topology is more complex than a dipolar one and that the present models did not
account for any evolution with time of the surface magnetic field. We obtain that the initial rotation of the models fitting $\sigma$ Ori E  is not much constrained and can be anywhere in the range studied here.
Because of its very young age, models predict no {significant} changes of the surface abundances due to rotational mixing for the main isotopes of the CNO elements as well as for helium. This young age remains compatible with an impact of the microscopic diffusion at least in zones stabilized by the surface magnetic field. In regions where low carbon abundance is obtained,
it would be very interesting to have nitrogen abundance determinations. If an increase of nitrogen would be found, this would support the view that some material processed by the CN cycle has been brought to the surface.


We see that the knowledge of both the surface velocity and of $P/\dot{P}$ is very constraining and allows to eliminate many models.
Also a change in metallicity, typically from 0.014 to 0.020 significantly changes the capacity of the models to provide a good fit.
It would be interesting to observe non-magnetic stars in the $\sigma$ Ori E cluster and to check whether indeed the metallicity is closer to 0.020 rather than 0.014.

It would be interesting to compute models tailored to sigma Ori E with different angular momentum transport.
For instance, if the stars rotate as solid bodies then the surface velocity can be maintained at a higher level everything else kept equal.
Indeed solid body rotation implies that the angular momentum is continuously transported from the core to the envelope\footnote{Local conservation of the angular momentum
produces a decrease of the angular velocity in the outer expanding layers and an increase in the contracting core, thus solid body rotation requires to slow down the core and accelerate the envelope.}. This implies,
at a given age, a higher surface velocity.
Now a higher surface velocity  also implies a stronger braking mechanism.
Thus models need to be computed to study the net effect. In addition, the chemical mixing is significantly changed and this has an impact on the evolutionary tracks, making it difficult to guess
what would be the result. This question will deserve a study on its own. 

\begin{acknowledgements}
The authors thanks the anonymous referee for her/his constructive report that has helped improving the paper.
They thank Georges Alecian for providing information on the microscopic diffusion timescales.
This work was sponsored by the
Swiss National Science Foundation (project number 200020-172505), National Natural Science Foundation
of China (grant No. 12173010),  Science and technology plan projects of Guizhou province  (Grant No. [2018]5781). 
GM, SE,  PE and CG have received funding from the European Research Council (ERC) under the European Union's Horizon 2020 research and innovation programme 
(grant agreement No 833925, project STAREX). GAW acknowledges support from the Discovery Grants program of the Natural Sciences and Engineering Research Council (NSERC) of Canada.
This work has made use of data from the European Space Agency (ESA) mission
{\it Gaia} (\url{https://www.cosmos.esa.int/gaia}), processed by the {\it Gaia}
Data Processing and Analysis Consortium (DPAC,
\url{https://www.cosmos.esa.int/web/gaia/dpac/consortium}). Funding for the DPAC
has been provided by national institutions, in particular the institutions
participating in the {\it Gaia} Multilateral Agreement.
\end{acknowledgements}

\bibliographystyle{aa}
\bibliography{sigmaorie}

\begin{thebibliography}{75}
\expandafter\ifx\csname natexlab\endcsname\relax\def\natexlab#1{#1}\fi

\bibitem[{{Babcock}(1947)}]{Babcock47}
{Babcock}, H.~W. 1947, \apj, 105, 105

\bibitem[{{Baume} {et~al.}(2003){Baume}, {V{\'a}zquez}, {Carraro}, \&
  {Feinstein}}]{Baume2003}
{Baume}, G., {V{\'a}zquez}, R.~A., {Carraro}, G., \& {Feinstein}, A. 2003,
  \aap, 402, 549

\bibitem[{{B{\'e}jar} {et~al.}(2001){B{\'e}jar}, {Mart{\'\i}n}, {Zapatero
  Osorio}, {Rebolo}, {Barrado y Navascu{\'e}s}, {Bailer-Jones}, {Mundt},
  {Baraffe}, {Chabrier}, \& {Allard}}]{Bejar01}
{B{\'e}jar}, V.~J.~S., {Mart{\'\i}n}, E.~L., {Zapatero Osorio}, M.~R., {et~al.}
  2001, \apj, 556, 830

\bibitem[{{Bell} {et~al.}(2013){Bell}, {Naylor}, {Mayne}, {Jeffries}, \&
  {Littlefair}}]{Bell2013}
{Bell}, C.~P.~M., {Naylor}, T., {Mayne}, N.~J., {Jeffries}, R.~D., \&
  {Littlefair}, S.~P. 2013, \mnras, 434, 806

\bibitem[{{Brott} {et~al.}(2011){Brott}, {de Mink}, {Cantiello}, {Langer}, {de
  Koter}, {Evans}, {Hunter}, {Trundle}, \& {Vink}}]{Brott2011}
{Brott}, I., {de Mink}, S.~E., {Cantiello}, M., {et~al.} 2011, \aap, 530, A115

\bibitem[{{Caballero}(2007)}]{Caba2007}
{Caballero}, J.~A. 2007, \aap, 466, 917

\bibitem[{{Caballero}(2017)}]{Caballero18NM}
{Caballero}, J.~A. 2017, Astronomische Nachrichten, 338, 629

\bibitem[{{Caballero}(2018)}]{Caballero18}
{Caballero}, J.~A. 2018, Research Notes of the American Astronomical Society,
  2, 25

\bibitem[{{Cunha} {et~al.}(1998){Cunha}, {Smith}, \& {Lambert}}]{Cunha1998}
{Cunha}, K., {Smith}, V.~V., \& {Lambert}, D.~L. 1998, \apj, 493, 195

\bibitem[{{de Jager} {et~al.}(1988){de Jager}, {Nieuwenhuijzen}, \& {van der
  Hucht}}]{deJager1988}
{de Jager}, C., {Nieuwenhuijzen}, H., \& {van der Hucht}, K.~A. 1988, \aaps,
  72, 281

\bibitem[{{Eikenberry} {et~al.}(2014){Eikenberry}, {Chojnowski}, {Wisniewski},
  {Majewski}, {Shetrone}, {Whelan}, {Bizyaev}, {Borish}, {Davenport}, {Ebelke},
  {Feuillet}, {Frinchaboy}, {Garner}, {Hearty}, {Holtzman}, {Li},
  {M{\'e}sz{\'a}ros}, {Nidever}, {Schneider}, {Skrutskie}, {Wilson}, \&
  {Zasowski}}]{Eikenberry2014}
{Eikenberry}, S.~S., {Chojnowski}, S.~D., {Wisniewski}, J., {et~al.} 2014,
  \apjl, 784, L30

\bibitem[{{Ekstr{\"o}m} {et~al.}(2012{\natexlab{a}}){Ekstr{\"o}m}, {Georgy},
  {Eggenberger}, {Meynet}, {Mowlavi}, {Wyttenbach}, {Granada}, {Decressin},
  {Hirschi}, {Frischknecht}, {Charbonnel}, \& {Maeder}}]{Ekstrom12}
{Ekstr{\"o}m}, S., {Georgy}, C., {Eggenberger}, P., {et~al.}
  2012{\natexlab{a}}, \aap, 537, A146

\bibitem[{{Ekstr{\"o}m} {et~al.}(2012{\natexlab{b}}){Ekstr{\"o}m}, {Georgy},
  {Granada}, {Wyttenbach}, \& {Meynet}}]{Ekstrom2012}
{Ekstr{\"o}m}, S., {Georgy}, C., {Granada}, A., {Wyttenbach}, A., \& {Meynet},
  G. 2012{\natexlab{b}}, ASPC, 453, 353

\bibitem[{{Ekstr{\"o}m} {et~al.}(2008){Ekstr{\"o}m}, {Meynet}, {Maeder}, \&
  {Barblan}}]{Ekstrom2008}
{Ekstr{\"o}m}, S., {Meynet}, G., {Maeder}, A., \& {Barblan}, F. 2008, \aap,
  478, 467

\bibitem[{{Espinosa Lara} \& {Rieutord}(2011)}]{Lara2011}
{Espinosa Lara}, F. \& {Rieutord}, M. 2011, \aap, 533, A43

\bibitem[{{Feiden}(2016)}]{Feid2016}
{Feiden}, G.~A. 2016, \aap, 593, A99

\bibitem[{{Fr{\'e}mat} {et~al.}(2005){Fr{\'e}mat}, {Zorec}, {Hubert}, \&
  {Floquet}}]{Fremat2005}
{Fr{\'e}mat}, Y., {Zorec}, J., {Hubert}, A.-M., \& {Floquet}, M. 2005, \aap,
  440, 305

\bibitem[{{Gaia Collaboration} {et~al.}(2018){Gaia Collaboration}, {Brown},
  {Vallenari}, {Prusti}, {de Bruijne}, {Babusiaux}, {Bailer-Jones}, {Biermann},
  {Evans}, {Eyer}, {Jansen}, {Jordi}, {Klioner}, {Lammers}, {Lindegren},
  {Luri}, {Mignard}, {Panem}, {Pourbaix}, {Randich}, {Sartoretti}, {Siddiqui},
  {Soubiran}, {van Leeuwen}, {Walton}, {Arenou}, {Bastian}, {Cropper},
  {Drimmel}, {Katz}, {Lattanzi}, {Bakker}, {Cacciari}, {Casta{\~n}eda},
  {Chaoul}, {Cheek}, {De Angeli}, {Fabricius}, {Guerra}, {Holl}, {Masana},
  {Messineo}, {Mowlavi}, {Nienartowicz}, {Panuzzo}, {Portell}, {Riello},
  {Seabroke}, {Tanga}, {Th{\'e}venin}, {Gracia-Abril}, {Comoretto},
  {Garcia-Reinaldos}, {Teyssier}, {Altmann}, {Andrae}, {Audard},
  {Bellas-Velidis}, {Benson}, {Berthier}, {Blomme}, {Burgess}, {Busso},
  {Carry}, {Cellino}, {Clementini}, {Clotet}, {Creevey}, {Davidson}, {De
  Ridder}, {Delchambre}, {Dell'Oro}, {Ducourant},
  {Fern{\'a}ndez-Hern{\'a}ndez}, {Fouesneau}, {Fr{\'e}mat}, {Galluccio},
  {Garc{\'\i}a-Torres}, {Gonz{\'a}lez-N{\'u}{\~n}ez}, {Gonz{\'a}lez-Vidal},
  {Gosset}, {Guy}, {Halbwachs}, {Hambly}, {Harrison}, {Hern{\'a}ndez},
  {Hestroffer}, {Hodgkin}, {Hutton}, {Jasniewicz}, {Jean-Antoine-Piccolo},
  {Jordan}, {Korn}, {Krone-Martins}, {Lanzafame}, {Lebzelter}, {L{\"o}ffler},
  {Manteiga}, {Marrese}, {Mart{\'\i}n-Fleitas}, {Moitinho}, {Mora}, {Muinonen},
  {Osinde}, {Pancino}, {Pauwels}, {Petit}, {Recio-Blanco}, {Richards},
  {Rimoldini}, {Robin}, {Sarro}, {Siopis}, {Smith}, {Sozzetti}, {S{\"u}veges},
  {Torra}, {van Reeven}, {Abbas}, {Abreu Aramburu}, {Accart}, {Aerts},
  {Altavilla}, {{\'A}lvarez}, {Alvarez}, {Alves}, {Anderson}, {Andrei},
  {Anglada Varela}, {Antiche}, {Antoja}, {Arcay}, {Astraatmadja}, {Bach},
  {Baker}, {Balaguer-N{\'u}{\~n}ez}, {Balm}, {Barache}, {Barata}, {Barbato},
  {Barblan}, {Barklem}, {Barrado}, {Barros}, {Barstow}, {Bartholom{\'e}
  Mu{\~n}oz}, {Bassilana}, {Becciani}, {Bellazzini}, {Berihuete}, {Bertone},
  {Bianchi}, {Bienaym{\'e}}, {Blanco-Cuaresma}, {Boch}, {Boeche}, {Bombrun},
  {Borrachero}, {Bossini}, {Bouquillon}, {Bourda}, {Bragaglia}, {Bramante},
  {Breddels}, {Bressan}, {Brouillet}, {Br{\"u}semeister}, {Brugaletta},
  {Bucciarelli}, {Burlacu}, {Busonero}, {Butkevich}, {Buzzi}, {Caffau},
  {Cancelliere}, {Cannizzaro}, {Cantat-Gaudin}, {Carballo}, {Carlucci},
  {Carrasco}, {Casamiquela}, {Castellani}, {Castro-Ginard}, {Charlot},
  {Chemin}, {Chiavassa}, {Cocozza}, {Costigan}, {Cowell}, {Crifo}, {Crosta},
  {Crowley}, {Cuypers}, {Dafonte}, {Damerdji}, {Dapergolas}, {David}, {David},
  {de Laverny}, {De Luise}, {De March}, {de Martino}, {de Souza}, {de Torres},
  {Debosscher}, {del Pozo}, {Delbo}, {Delgado}, {Delgado}, {Di Matteo},
  {Diakite}, {Diener}, {Distefano}, {Dolding}, {Drazinos}, {Dur{\'a}n},
  {Edvardsson}, {Enke}, {Eriksson}, {Esquej}, {Eynard Bontemps}, {Fabre},
  {Fabrizio}, {Faigler}, {Falc{\~a}o}, {Farr{\`a}s Casas}, {Federici},
  {Fedorets}, {Fernique}, {Figueras}, {Filippi}, {Findeisen}, {Fonti},
  {Fraile}, {Fraser}, {Fr{\'e}zouls}, {Gai}, {Galleti}, {Garabato},
  {Garc{\'\i}a-Sedano}, {Garofalo}, {Garralda}, {Gavel}, {Gavras}, {Gerssen},
  {Geyer}, {Giacobbe}, {Gilmore}, {Girona}, {Giuffrida}, {Glass}, {Gomes},
  {Granvik}, {Gueguen}, {Guerrier}, {Guiraud}, {Guti{\'e}rrez-S{\'a}nchez},
  {Haigron}, {Hatzidimitriou}, {Hauser}, {Haywood}, {Heiter}, {Helmi}, {Heu},
  {Hilger}, {Hobbs}, {Hofmann}, {Holland}, {Huckle}, {Hypki}, {Icardi},
  {Jan{\ss}en}, {Jevardat de Fombelle}, {Jonker}, {Juh{\'a}sz}, {Julbe},
  {Karampelas}, {Kewley}, {Klar}, {Kochoska}, {Kohley}, {Kolenberg},
  {Kontizas}, {Kontizas}, {Koposov}, {Kordopatis}, {Kostrzewa-Rutkowska},
  {Koubsky}, {Lambert}, {Lanza}, {Lasne}, {Lavigne}, {Le Fustec}, {Le
  Poncin-Lafitte}, {Lebreton}, {Leccia}, {Leclerc}, {Lecoeur-Taibi},
  {Lenhardt}, {Leroux}, {Liao}, {Licata}, {Lindstr{\o}m}, {Lister}, {Livanou},
  {Lobel}, {L{\'o}pez}, {Managau}, {Mann}, {Mantelet}, {Marchal}, {Marchant},
  {Marconi}, {Marinoni}, {Marschalk{\'o}}, {Marshall}, {Martino}, {Marton},
  {Mary}, {Massari}, {Matijevi{\v{c}}}, {Mazeh}, {McMillan}, {Messina},
  {Michalik}, {Millar}, {Molina}, {Molinaro}, {Moln{\'a}r}, {Montegriffo},
  {Mor}, {Morbidelli}, {Morel}, {Morris}, {Mulone}, {Muraveva}, {Musella},
  {Nelemans}, {Nicastro}, {Noval}, {O'Mullane}, {Ord{\'e}novic},
  {Ord{\'o}{\~n}ez-Blanco}, {Osborne}, {Pagani}, {Pagano}, {Pailler},
  {Palacin}, {Palaversa}, {Panahi}, {Pawlak}, {Piersimoni}, {Pineau}, {Plachy},
  {Plum}, {Poggio}, {Poujoulet}, {Pr{\v{s}}a}, {Pulone}, {Racero}, {Ragaini},
  {Rambaux}, {Ramos-Lerate}, {Regibo}, {Reyl{\'e}}, {Riclet}, {Ripepi}, {Riva},
  {Rivard}, {Rixon}, {Roegiers}, {Roelens}, {Romero-G{\'o}mez}, {Rowell},
  {Royer}, {Ruiz-Dern}, {Sadowski}, {Sagrist{\`a} Sell{\'e}s}, {Sahlmann},
  {Salgado}, {Salguero}, {Sanna}, {Santana-Ros}, {Sarasso}, {Savietto},
  {Schultheis}, {Sciacca}, {Segol}, {Segovia}, {S{\'e}gransan}, {Shih},
  {Siltala}, {Silva}, {Smart}, {Smith}, {Solano}, {Solitro}, {Sordo}, {Soria
  Nieto}, {Souchay}, {Spagna}, {Spoto}, {Stampa}, {Steele},
  {Steidelm{\"u}ller}, {Stephenson}, {Stoev}, {Suess}, {Surdej}, {Szabados},
  {Szegedi-Elek}, {Tapiador}, {Taris}, {Tauran}, {Taylor}, {Teixeira},
  {Terrett}, {Teyssand ier}, {Thuillot}, {Titarenko}, {Torra Clotet}, {Turon},
  {Ulla}, {Utrilla}, {Uzzi}, {Vaillant}, {Valentini}, {Valette}, {van Elteren},
  {Van Hemelryck}, {van Leeuwen}, {Vaschetto}, {Vecchiato}, {Veljanoski},
  {Viala}, {Vicente}, {Vogt}, {von Essen}, {Voss}, {Votruba}, {Voutsinas},
  {Walmsley}, {Weiler}, {Wertz}, {Wevers}, {Wyrzykowski}, {Yoldas},
  {{\v{Z}}erjal}, {Ziaeepour}, {Zorec}, {Zschocke}, {Zucker}, {Zurbach}, \&
  {Zwitter}}]{DR22018}
{Gaia Collaboration}, {Brown}, A.~G.~A., {Vallenari}, A., {et~al.} 2018, \aap,
  616, A1

\bibitem[{{Georgy} {et~al.}(2013){Georgy}, {Ekstr{\"o}m}, {Eggenberger},
  {Meynet}, {Haemmerl{\'e}}, {Maeder}, {Granada}, {Groh}, {Hirschi}, {Mowlavi},
  {Yusof}, {Charbonnel}, {Decressin}, \& {Barblan}}]{Georgy13a}
{Georgy}, C., {Ekstr{\"o}m}, S., {Eggenberger}, P., {et~al.} 2013, \aap, 558,
  A103

\bibitem[{{Georgy} {et~al.}(2014){Georgy}, {Granada}, {Ekstr{\"o}m}, {Meynet},
  {Anderson}, {Wyttenbach}, {Eggenberger}, \& {Maeder}}]{Georgy2014}
{Georgy}, C., {Granada}, A., {Ekstr{\"o}m}, S., {et~al.} 2014, \aap, 566, A21

\bibitem[{{Georgy} {et~al.}(2017){Georgy}, {Meynet}, {Ekstr{\"o}m}, {Wade},
  {Petit}, {Keszthelyi}, \& {Hirschi}}]{Georgy2017}
{Georgy}, C., {Meynet}, G., {Ekstr{\"o}m}, S., {et~al.} 2017, \aap, 599, L5

\bibitem[{{Groote} \& {Hunger}(1982)}]{GH1982}
{Groote}, D. \& {Hunger}, K. 1982, \aap, 116, 64

\bibitem[{{Grunhut} {et~al.}(2012){Grunhut}, {Rivinius}, {Wade}, {Townsend},
  {Marcolino}, {Bohlender}, {Szeifert}, {Petit}, {Matthews}, {Rowe}, {Moffat},
  {Kallinger}, {Kuschnig}, {Guenther}, {Rucinski}, {Sasselov}, \&
  {Weiss}}]{Grunhut2012}
{Grunhut}, J.~H., {Rivinius}, T., {Wade}, G.~A., {et~al.} 2012, \mnras, 419,
  1610

\bibitem[{{Haemmerl{\'e}} {et~al.}(2019){Haemmerl{\'e}}, {Eggenberger},
  {Ekstr{\"o}m}, {Georgy}, {Meynet}, {Postel}, {Audard}, {S{\o}rensen}, \&
  {Fragos}}]{Haemerle2019}
{Haemmerl{\'e}}, L., {Eggenberger}, P., {Ekstr{\"o}m}, S., {et~al.} 2019, \aap,
  624, A137

\bibitem[{{Herbig}(1998)}]{Herbig1998}
{Herbig}, G.~H. 1998, \apj, 497, 736

\bibitem[{{Hern{\'a}ndez} {et~al.}(2005){Hern{\'a}ndez}, {Calvet}, {Hartmann},
  {Brice{\~n}o}, {Sicilia-Aguilar}, \& {Berlind}}]{Her2005}
{Hern{\'a}ndez}, J., {Calvet}, N., {Hartmann}, L., {et~al.} 2005, \aj, 129, 856

\bibitem[{{Hubrig} {et~al.}(2015){Hubrig}, {Sch{\"o}ller}, {Fossati}, {Morel},
  {Castro}, {Oskinova}, {Przybilla}, {Eikenberry}, {Nieva}, \&
  {Langer}}]{Hubrig2015}
{Hubrig}, S., {Sch{\"o}ller}, M., {Fossati}, L., {et~al.} 2015, \aap, 578, L3

\bibitem[{{Hunger} {et~al.}(1989){Hunger}, {Heber}, \& {Groote}}]{Hunger1989}
{Hunger}, K., {Heber}, U., \& {Groote}, D. 1989, \aap, 224, 57

\bibitem[{{Jerzykiewicz}(1993)}]{Jery1993}
{Jerzykiewicz}, M. 1993, \aaps, 97, 421

\bibitem[{{Keszthelyi} {et~al.}(2020){Keszthelyi}, { Shultz}, { David-Uraz},
  {ud-Doula}, { Townsend}, {Wade}, \& {Georgy}}]{Zsolt2020}
{Keszthelyi}, Z.~{Meynet}, G., { Shultz}, M.~E., { David-Uraz}, A., {et~al.}
  2020, \mnras, 493, 518

\bibitem[{{Keszthelyi} {et~al.}(2019){Keszthelyi}, {Georgy}, {Wade}, {Petit},
  \& {David-Uraz}}]{Zsolt2019}
{Keszthelyi}, Z.~{Meynet}, G., {Georgy}, C., {Wade}, G.~A., {Petit}, V., \&
  {David-Uraz}, A. 2019, \mnras, 485, 5843

\bibitem[{{Keszthelyi} \& {Petit}(2017)}]{Zsolt2017}
{Keszthelyi}, Z.~and{ Wade}, G.~A. \& {Petit}, V. 2017, IAUS, 329, 250

\bibitem[{{Keszthelyi} {et~al.}(2021){Keszthelyi}, {Meynet}, {Martins}, {de
  Koter}, \& {David-Uraz}}]{Zsolt2021}
{Keszthelyi}, Z., {Meynet}, G., {Martins}, F., {de Koter}, A., \& {David-Uraz},
  A. 2021, \mnras

\bibitem[{{Krti{\v{c}}ka} {et~al.}(2020){Krti{\v{c}}ka}, {Mikul{\'a}{\v{s}}ek},
  {Prv{\'a}k}, {Niemczura}, {Leone}, \& {Wade}}]{Krticka2020}
{Krti{\v{c}}ka}, J., {Mikul{\'a}{\v{s}}ek}, Z., {Prv{\'a}k}, M., {et~al.} 2020,
  \mnras, 493, 2140

\bibitem[{{Landstreet} \& {Borra}(1978)}]{Landstreet1978}
{Landstreet}, J.~D. \& {Borra}, E.~F. 1978, \apjl, 224, L5

\bibitem[{{Lindegren} {et~al.}(2020){Lindegren}, {Klioner}, {Hern{\'a}ndez},
  {Bombrun}, {Ramos-Lerate}, {Steidelm{\"u}ller}, {Bastian}, {Biermann}, {de
  Torres}, {Gerlach}, {Geyer}, {Hilger}, {Hobbs}, {Lammers}, {McMillan},
  {Stephenson}, {Casta{\~n}eda}, {Davidson}, {Fabricius}, {Gracia-Abril},
  {Portell}, {Rowell}, {Teyssier}, {Torra}, {Bartolom{\'e}}, {Clotet},
  {Garralda}, {Gonz{\'a}lez-Vidal}, {Torra}, {Abbas}, {Altmann}, {Anglada
  Varela}, {Balaguer-N{\'u}{\~n}ez}, {Balog}, {Barache}, {Becciani}, {Bernet},
  {Bertone}, {Bianchi}, {Bouquillon}, {Brown}, {Bucciarelli}, {Busonero},
  {Butkevich}, {Buzzi}, {Cancelliere}, {Carlucci}, {Charlot}, {Cioni},
  {Crosta}, {Crowley}, {del Peloso}, {del Pozo}, {Drimmel}, {Esquej}, {Fienga},
  {Fraile}, {Gai}, {Garcia-Reinaldos}, {Guerra}, {Hambly}, {Hauser},
  {Jan{\ss}en}, {Jordan}, {Kostrzewa-Rutkowska}, {Lattanzi}, {Liao}, {Licata},
  {Lister}, {L{\"o}ffler}, {Marchant}, {Masip}, {Mignard}, {Mints}, {Molina},
  {Mora}, {Morbidelli}, {Murphy}, {Pagani}, {Panuzzo}, {Pe{\~n}alosa Esteller},
  {Poggio}, {Re Fiorentin}, {Riva}, {Sagrist{\`a} Sell{\'e}s}, {Sanchez
  Gimenez}, {Sarasso}, {Sciacca}, {Siddiqui}, {Smart}, {Souami}, {Spagna},
  {Steele}, {Taris}, {Utrilla}, {van Reeven}, \&
  {Vecchiato}}]{Lindegren_etal20}
{Lindegren}, L., {Klioner}, S.~A., {Hern{\'a}ndez}, J., {et~al.} 2020, arXiv
  e-prints, arXiv:2012.03380

\bibitem[{{Luri} {et~al.}(2018){Luri}, {Brown}, {Sarro}, {Arenou},
  {Bailer-Jones}, {Castro-Ginard}, {de Bruijne}, {Prusti}, {Babusiaux}, \&
  {Delgado}}]{Luri_etal18}
{Luri}, X., {Brown}, A.~G.~A., {Sarro}, L.~M., {et~al.} 2018, \aap, 616, A9

\bibitem[{{Marigo} {et~al.}(2008){Marigo}, {Girardi}, {Bressan}, {Groenewegen},
  {Silva}, \& {Granato}}]{Mar2008}
{Marigo}, P., {Girardi}, L., {Bressan}, A., {et~al.} 2008, \aap, 482, 883

\bibitem[{{Meynet} {et~al.}(2011){Meynet}, {Eggenberger}, \&
  {Maeder}}]{Meynet11}
{Meynet}, G., {Eggenberger}, P., \& {Maeder}, A. 2011, \aap, 525, L11

\bibitem[{{Michaud}(1970)}]{Michaud1970}
{Michaud}, G. 1970, \apj, 160, 641

\bibitem[{{Michaud} {et~al.}(2015){Michaud}, {Alecian}, \&
  {Richer}}]{Michaud2015}
{Michaud}, G., {Alecian}, G., \& {Richer}, J. 2015, {Atomic Diffusion in Stars}

\bibitem[{{Mikul{\'a}{\v s}ek} {et~al.}(2008){Mikul{\'a}{\v s}ek}, {Krti{\v
  c}ka}, {Henry}, {de Villiers}, {Paunzen}, \& {Zejda}}]{Miku2008}
{Mikul{\'a}{\v s}ek}, Z., {Krti{\v c}ka}, J., {Henry}, G.~W., {et~al.} 2008,
  \aap, 485, 585

\bibitem[{{Mikul{\'a}{\v s}ek} {et~al.}(2010){Mikul{\'a}{\v s}ek}, {Krti{\v
  c}ka}, {Henry}, {de Villiers}, {Paunzen}, \& {Zejda}}]{Miku2010}
{Mikul{\'a}{\v s}ek}, Z., {Krti{\v c}ka}, J., {Henry}, G.~W., {et~al.} 2010,
  \aap, 511, L7

\bibitem[{{Mikul{\'a}{\v{s}}ek} {et~al.}(2011){Mikul{\'a}{\v{s}}ek},
  {Krti{\v{c}}ka}, {Henry}, {Jan{\'\i}k}, {Zverko},
  {{\v{Z}}i{\v{z}}{\v{n}}ovsk{\'y}}, {Zejda}, {Li{\v{s}}ka},
  {Zv{\v{e}}{\v{r}}ina}, {Kudrjavtsev}, {Romanyuk}, {Sokolov}, {L{\"u}ftinger},
  {Trigilio}, {Neiner}, \& {de Villiers}}]{Mikul2011}
{Mikul{\'a}{\v{s}}ek}, Z., {Krti{\v{c}}ka}, J., {Henry}, G.~W., {et~al.} 2011,
  \aap, 534, L5

\bibitem[{{Oksala} {et~al.}(2015){Oksala}, {Kochukhov}, {Krti{\v{c}}ka},
  {Townsend}, {Wade}, {Prv{\'a}k}, {Mikul{\'a}{\v{s}}ek}, {Silvester}, \&
  {Owocki}}]{Oksala2015}
{Oksala}, M.~E., {Kochukhov}, O., {Krti{\v{c}}ka}, J., {et~al.} 2015, \mnras,
  451, 2015

\bibitem[{{Oksala} {et~al.}(2010){Oksala}, {Wade}, {Marcolino}, {Grunhut},
  {Bohlender}, {Manset}, {Townsend}, \& {Mimes Collaboration}}]{Ok2010}
{Oksala}, M.~E., {Wade}, G.~A., {Marcolino}, W.~L.~F., {et~al.} 2010, \mnras,
  405, L51

\bibitem[{{Oksala} {et~al.}(2012){Oksala}, {Wade}, {Townsend}, {Owocki},
  {Kochukhov}, {Neiner}, {Alecian}, \& {Grunhut}}]{Oksala2012}
{Oksala}, M.~E., {Wade}, G.~A., {Townsend}, R.~H.~D., {et~al.} 2012, \mnras,
  419, 959

\bibitem[{{Oliveira} {et~al.}(2002){Oliveira}, {Jeffries}, {Kenyon},
  {Thompson}, \& {Naylor}}]{Oliveira02}
{Oliveira}, J.~M., {Jeffries}, R.~D., {Kenyon}, M.~J., {Thompson}, S.~A., \&
  {Naylor}, T. 2002, \aap, 382, L22

\bibitem[{{Osmer} \& {Peterson}(1974)}]{Osmer1974}
{Osmer}, P.~S. \& {Peterson}, D.~M. 1974, \apj, 187, 117

\bibitem[{{Panei} {et~al.}(2021){Panei}, {Vallverd{\'u}}, \&
  {Cidale}}]{Panei2021}
{Panei}, J.~A., {Vallverd{\'u}}, R.~E., \& {Cidale}, L.~S. 2021, \aap, 650, A92

\bibitem[{{Petit} {et~al.}(2017){Petit}, {Keszthelyi}, {MacInnis}, {Cohen},
  {Townsend}, {Wade}, {Thomas}, {Owocki}, {Puls}, \& {ud-Doula}}]{Petit2017}
{Petit}, V., {Keszthelyi}, Z., {MacInnis}, R., {et~al.} 2017, \mnras, 466, 1052

\bibitem[{{Petit} {et~al.}(2013){Petit}, {Owocki}, {Wade}, {Cohen},
  {Sundqvist}, {Gagn{\'e}}, , {Oksala}, {Bohlender}, {Rivinius}, {Henrichs},
  {Alecian}, {Townsend}, {ud-Doula}, \& Collaboration}]{Petit2013}
{Petit}, V., {Owocki}, S.~P., {Wade}, G.~A., {et~al.} 2013, \mnras, 429, 398

\bibitem[{{Przybilla} {et~al.}(2016){Przybilla}, {Fossati}, {Hubrig}, {Nieva},
  {J{\"a}rvinen}, {Castro}, {Sch{\"o}ller}, {Ilyin}, {Butler}, {Schneider},
  {Oskinova}, {Morel}, {Langer}, {de Koter}, \& {BOB Collaboration}}]{Przy2016}
{Przybilla}, N., {Fossati}, L., {Hubrig}, S., {et~al.} 2016, \aap, 587, A7

\bibitem[{{Reiners} {et~al.}(2000){Reiners}, {Stahl}, {Wolf}, {Kaufer}, \&
  {Rivinius}}]{Reiners2000}
{Reiners}, A., {Stahl}, O., {Wolf}, B., {Kaufer}, A., \& {Rivinius}, T. 2000,
  \aap, 363, 585

\bibitem[{{Rivinius} {et~al.}(2008){Rivinius}, {Tefl}, {Townsend}, \&
  {Baade}}]{Rivi2008}
{Rivinius}, T., {Tefl}, S.~{\AA}., {Townsend}, R.~H.~D., \& {Baade}, D. 2008,
  \aap, 482, 255

\bibitem[{{Rivinius} {et~al.}(2013){Rivinius}, {Townsend}, {Kochukhov}, {{\v
  S}tefl}, {Baade}, {Barrera}, \& {Szeifert}}]{Rivi2013}
{Rivinius}, T., {Townsend}, R.~H.~D., {Kochukhov}, O., {et~al.} 2013, \mnras,
  429, 177

\bibitem[{{Schaefer} {et~al.}(2016){Schaefer}, {Hummel}, {Gies}, {Zavala},
  {Monnier}, {Walter}, {Turner}, {Baron}, {ten Brummelaar}, {Che},
  {Farrington}, {Kraus}, {Sturmann}, \& {Sturmann}}]{SchaeferHummelGies_etal16}
{Schaefer}, G.~H., {Hummel}, C.~A., {Gies}, D.~R., {et~al.} 2016, \aj, 152, 213

\bibitem[{{Scholz} {et~al.}(1999){Scholz}, {Brunzendorf}, {Ivanov},
  {Kharchenko}, {Lasker}, {Meusinger}, {Preibisch}, {Schilbach}, \&
  {Zinnecker}}]{Scholz1999}
{Scholz}, R.-D., {Brunzendorf}, J., {Ivanov}, G., {et~al.} 1999, \aaps, 137,
  305

\bibitem[{{Scott} {et~al.}(2015){Scott}, {Asplund}, {Grevesse}, {Bergemann}, \&
  {Sauval}}]{Scott2015}
{Scott}, P., {Asplund}, M., {Grevesse}, N., {Bergemann}, M., \& {Sauval}, A.~J.
  2015, \aap, 573, A26

\bibitem[{{Sherry} {et~al.}(2008){Sherry}, {Walter}, {Wolk}, \&
  {Adams}}]{Sherry2008}
{Sherry}, W.~H., {Walter}, F.~M., {Wolk}, S.~J., \& {Adams}, N.~R. 2008, \aj,
  135, 1616

\bibitem[{{Shultz} {et~al.}(2019{\natexlab{a}}){Shultz}, {Rivinius}, {Das},
  {Wade}, \& {Chandra}}]{Shultz2019b}
{Shultz}, M., {Rivinius}, T., {Das}, B., {Wade}, G.~A., \& {Chandra}, P.
  2019{\natexlab{a}}, \mnras, 486, 5558

\bibitem[{{Shultz} {et~al.}(2020){Shultz}, {Owocki}, {Rivinius}, {Wade},
  {Neiner}, {Alecian}, {Kochukhov}, {Bohlender}, {ud-Doula}, {Landstreet},
  {Sikora}, {David-Uraz}, {Petit}, {Cerraho{\u{g}}lu}, {Fine}, {Henson}, {MiMeS
  Collaboration}, \& {BinaMIcS Collaboration}}]{Shultz2020}
{Shultz}, M.~E., {Owocki}, S., {Rivinius}, T., {et~al.} 2020, \mnras, 499, 5379

\bibitem[{{Shultz} {et~al.}(2019{\natexlab{b}}){Shultz}, {Wade}, {Rivinius},
  {Alecian}, {Neiner}, {Petit}, {Wisniewski}, {MiMeS Collaboration}, \&
  {BinaMIcS Collaboration}}]{Shultz2019a}
{Shultz}, M.~E., {Wade}, G.~A., {Rivinius}, T., {et~al.} 2019{\natexlab{b}},
  \mnras, 485, 1508

\bibitem[{{Sikora} {et~al.}(2015){Sikora}, {Wade}, {Bohlender}, {Neiner},
  {Oksala}, {Shultz}, {Cohen}, {ud-Doula}, {Grunhut}, {Monin}, {Owocki},
  {Petit}, {Rivinus}, \& {Townsend}}]{Sikora2015}
{Sikora}, J., {Wade}, G.~A., {Bohlender}, D.~A., {et~al.} 2015, \mnras, 451,
  1928

\bibitem[{{Stift} \& {Alecian}(2016)}]{Stift2016}
{Stift}, M.~J. \& {Alecian}, G. 2016, \mnras, 457, 74

\bibitem[{{Townsend} {et~al.}(2010){Townsend}, {Oksala}, {Cohen}, {Owocki}, \&
  {ud-Doula}}]{Townsend2010}
{Townsend}, R.~H.~D., {Oksala}, M.~E., {Cohen}, D.~H., {Owocki}, S.~P., \&
  {ud-Doula}, A. 2010, \apjl, 714, L318

\bibitem[{{Townsend} \& {Owocki}(2005)}]{Townsend2005}
{Townsend}, R.~H.~D. \& {Owocki}, S.~P. 2005, \mnras, 357, 251

\bibitem[{{Townsend} {et~al.}(2005){Townsend}, {Owocki}, \& {Groote}}]{T05}
{Townsend}, R.~H.~D., {Owocki}, S.~P., \& {Groote}, D. 2005, \apjl, 630, L81

\bibitem[{{Townsend} {et~al.}(2013){Townsend}, {Rivinius}, {Rowe}, {Moffat},
  {Matthews}, {Bohlender}, {Neiner}, {Telting}, {Guenther}, {Kallinger},
  {Kuschnig}, {Rucinski}, {Sasselov}, \& {Weiss}}]{Townsend2013}
{Townsend}, R.~H.~D., {Rivinius}, T., {Rowe}, J.~F., {et~al.} 2013, \apj, 769,
  33

\bibitem[{{ud-Doula} \& {Owocki}(2002)}]{Ud-Doula2002}
{ud-Doula}, A. \& {Owocki}, S.~P. 2002, \apj, 576, 413

\bibitem[{{ud-Doula} {et~al.}(2009){ud-Doula}, {Owocki}, \&
  {Townsend}}]{Ud2009}
{ud-Doula}, A., {Owocki}, S.~P., \& {Townsend}, R.~H.~D. 2009, \mnras, 392,
  1022

\bibitem[{{Ud-Doula} {et~al.}(2009){Ud-Doula}, {Owocki}, \&
  {Townsend}}]{udDoula2009}
{Ud-Doula}, A., {Owocki}, S.~P., \& {Townsend}, R. H.~D. 2009, \mnras, 392,
  1022

\bibitem[{{von Zeipel}(1924)}]{Zeip1924}
{von Zeipel}, H. 1924, \mnras, 84, 665

\bibitem[{{Wolff}(1968)}]{wolff68}
{Wolff}, S.~C. 1968, \pasp, 80, 281

\bibitem[{{Zahn}(1992)}]{Zahn1992}
{Zahn}, J.-P. 1992, \aap, 265, 115

\end{thebibliography}
\end{document}